\documentclass[12pt, twocolumn]{article}
\usepackage[utf8]{inputenc}
\usepackage{graphicx} 
\usepackage{amsmath}
\usepackage[letterpaper, portrait, margin=1in]{geometry}
\usepackage[nottoc]{tocbibind} 

\usepackage{amssymb}
\usepackage[labelfont=bf]{caption}
\usepackage{subfig}
\usepackage{setspace}
\usepackage{titlesec}
\usepackage{wrapfig}
\usepackage{float}
\usepackage{cite}
\usepackage{url}

\newcommand{\beginsupplement}{
    \setcounter{section}{0}
    \renewcommand{\thesection}{S\arabic{section}}
    \setcounter{equation}{0}
    \renewcommand{\theequation}{S\arabic{equation}}
    \setcounter{figure}{0}
    \renewcommand{\thefigure}{S\arabic{figure}}
    \setcounter{table}{0}
    \renewcommand{\thetable}{S\arabic{table}}
    \newcounter{offset}
  \setcounter{offset}{\value{figure}}
  \renewcommand{\thefigure}{S\the\numexpr\value{figure}-\value{offset}\relax}
}

\usepackage{hyperref}
\usepackage{blindtext}

\titleformat{\section}[block]
  {\normalfont\large\bfseries}{\thesection}{1em}{\large}

\title{Redundant contacts and force redistribution stabilize limbless vertical climbing}  

\author{Calvin A. Riiska$^{1}$, Michelle Lee$^{1}$, Yonatan Nemenman$^{1}$ \\ Gauge Thacker$^{1}$, Joseph R. Mendelson III$^{2,3}$,  Jennifer M. Rieser$^{1}$\\
\small $^{1}$Emory University Department of Physics, Atlanta, GA\\
\small$^{2}$Department of Research, Zoo Atlanta, Atlanta, GA\\
\small$^{3}$School of Biological Sciences, Georgia Institute of Technology, Atlanta, GA\\}
\date{\today }

\begin{document}

\maketitle

\begin{abstract}
Animals navigating complex vertical environments must secure stable footholds, a challenge for species without feet. While arboreal climbing has evolved repeatedly in snakes, the physical mechanisms they use to scale broad, nearly flat surfaces remain poorly understood. By measuring three-dimensional body kinematics and per-contact forces on a smooth vertical wall with protruding posts, we show that cornsnakes climb by dynamically balancing forces across a highly redundant network of 5 to 16 simultaneous contacts—far exceeding the three contacts minimally required for physical stability. Using a computational model and a robotic climber, we demonstrate that while simple body undulations and passive friction are mechanically sufficient to climb this terrain, snakes systematically deviate from this passive baseline. While downward climbing relies primarily on friction, ascending snakes actively generate positive mechanical work at their contacts to propel themselves. Furthermore, we found that whenever a snake engages a new contact, it triggers a stereotyped, system-wide redistribution of force that seamlessly integrates the new foothold without disrupting whole-body balance. These results reveal how a continuous, flexible body can transform sparse environmental features into a robust, fault-tolerant network. This mechanism provides a biomechanical framework for understanding the repeated evolution of limbless climbing and offers physical principles for designing agile robots for unstructured terrain.
\end{abstract}

\section*{Introduction}

Arboreal habitats impose extreme physical challenges, where animals must negotiate contacts with branches and tree trunks of variable texture, size, shape, and flexibility, where failure can lead to catastrophic falls. In limbed animals, surface contacts are often established with adapted traits including claws, adhesive pads, grasping limbs, and prehensile tails to meet these demands~\cite{labonte_extreme_2016,autumn2000adhesive,dirks_fluid-based_2011,cant_positional_1992,fischer_evolution_2010,hunt_acrobatic_2021,jusufi_active_2008}. Snakes, however, present a striking contrast. Despite the lack of limbs and specialized attachment structures, climbing has evolved repeatedly in snakes~\cite{pizzatto2007life}. Many species are capable of traversing inclined or vertical environments, ranging from narrow branches and dense vegetation to tree trunks, rock faces, and walls~\cite{astley_effects_2007,astley_arboreal_2009,jayne_what_2020,byrnes_gripping_2014,savidge_lasso_2021}, where contact  geometry and availability vary dramatically. How a continuous, limbless body generates support, stability, and propulsion across such diverse terrain remains poorly understood. 

Previous studies have primarily examined environments that allow snakes to wrap around a substrate, brace between opposing surfaces, or push against regularly distributed obstacles. On thin branches and tree trunks, snakes commonly climb upward using concertina locomotion, involving cyclic anchoring and advancing body segments, and constrict the surface to support their weight~\cite{jayne_what_2020,astley_effects_2007,astley_arboreal_2009,byrnes_gripping_2014,savidge_lasso_2021}. With anchored regions supporting the animal’s weight, other regions advance or extend across gaps while maintaining posture~\cite{jurestovsky_generation_2021,jayne_surface_2012,jorgensen_three-dimensional_2017,hoffmann2026postural}. In confined, inclined channels, snakes instead brace against opposing walls to increase normal forces and thereby generate the friction needed for support~\cite{marvi_friction_2012}. Surface protrusions on poles or within channels allow a shift in gait to lateral undulation, involving waves propagating down the body, by providing discrete contact sites to push against. Lateral undulation is favorable in that it is substantially faster~\cite{jayne_why_2015,astley_arboreal_2009} and presumed to be more energetically efficient than concertina~\cite{jayne_muscular_1988,jayne_muscular_1988-1,jayne1986kinematics}.

Many surfaces encountered by snakes are flat or nearly flat such as large tree trunks, rocky surfaces, and brick walls where constriction and bracing between opposing sides becomes impossible. That snakes climb such surfaces at all therefore poses a basic mechanical puzzle: how can a limbless body use discrete, sometimes sparse asperities to 
support, stabilize, and propel itself during vertical ascent and descent? Friction remains essential, and some species exhibit remarkable control over their ventral scales, flaring them and forming keels that grip or wedge against surface features~\cite{marvi_friction_2012,jayne_why_2015}. Yet not all species can climb~\cite{lillywhite2014snakes,tingle2024functional}, indicating that the limbless body plan provides an opportunity rather than a guarantee. 

Understanding how snakes exploit discrete surface features therefore requires treating climbing as a dynamic multi-contact problem. Unlike limbed animals, whose contacts are generally associated with a fixed set of appendages, snakes can establish contacts at many locations along a continuous, deformable body. In limbless locomotion, cycles of limb placement are replaced by transient body--substrate connections that form and propagate along the body as it deforms~\cite{rieser_geometric_2024}, allowing forces to be applied at multiple locations simultaneously. This distributed contact network provides substantial mechanical redundancy, but how snakes redistribute forces as contacts are gained, lost, and repositioned remains largely unknown.
 
In this study, we recorded climbing behavior in cornsnakes (\textit{Pantherophis guttatus}) ascending and descending a flat, vertical wall instrumented with force sensitive posts, combining kinematic marker tracking data with time-resolved measurements of the forces at each contact. We found climbing to be quasi-static: net force and torque remained close to balance, and accelerations were small throughout each trial. The contact network was also highly redundant, with snakes maintaining far more contacts than planar force and torque balance requires. A minimal computational model showed that a prescribed body wave with entirely passive contact mechanics was sufficient for climbing, and an open-loop-controlled robot demonstrated the physical viability of this strategy. Snake descents closely matched this passive prediction and were dominated by dissipative contacts. During ascent, however,  snakes actively redirected contact forces beyond the predictions for passive Coulomb friction. Although overall force patterns changed when posts were shortened or spaced further apart, snakes used similar stereotyped, balance-preserving force rearrangements across conditions. Together, these results reveal how a continuous, limbless body can climb surfaces that cannot be continuously wrapped or gripped, while providing a framework for comparing climbing strategies across substrates and designing limbless robots for vertical, complex terrain.

\section*{Results and Discussion}

\begin{figure*}[ht!]
	\includegraphics[width=\textwidth]{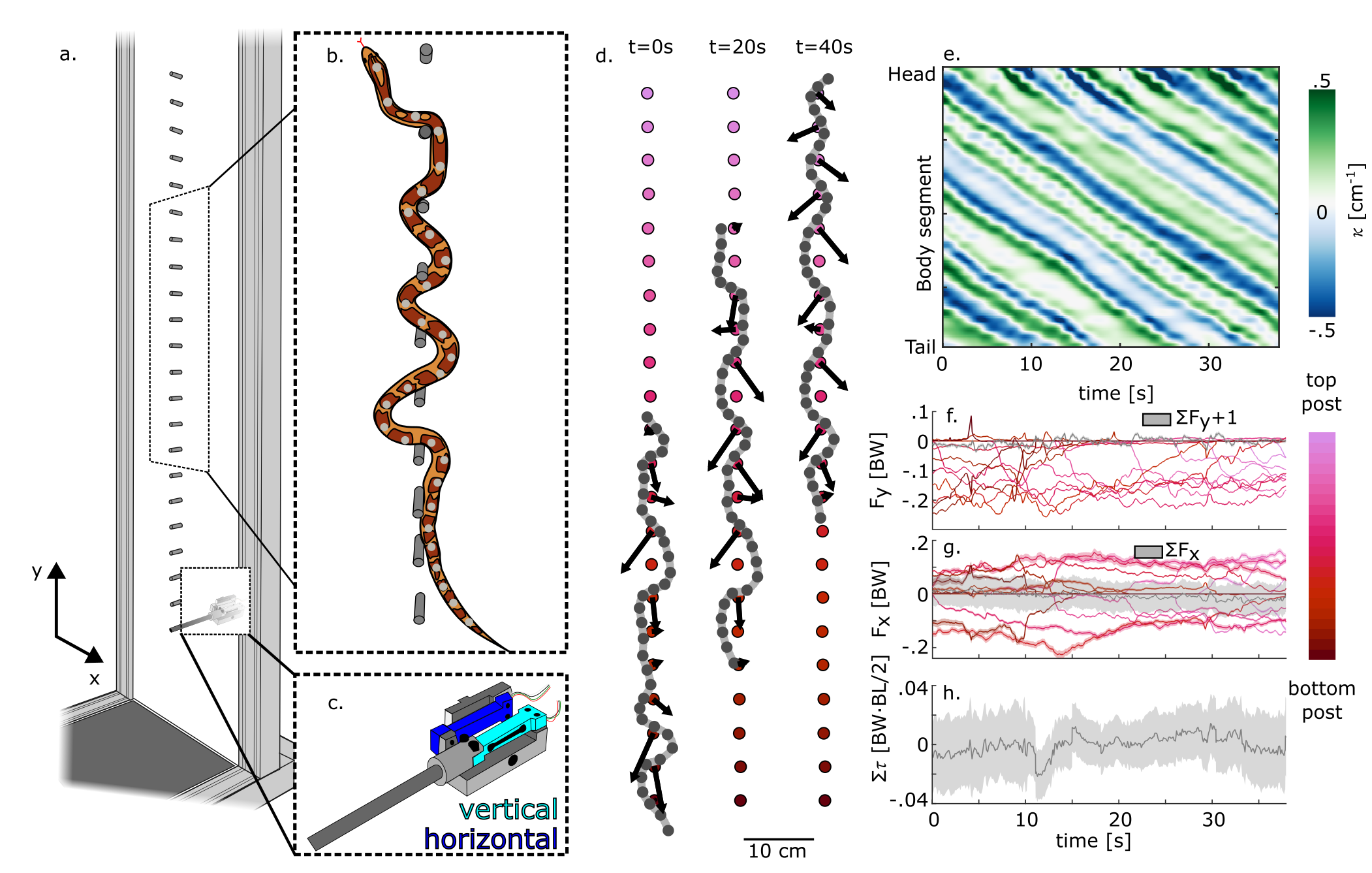}
\caption{\textbf{A force-sensing wall resolves body kinematics and per-post forces throughout climbing.} \textbf{(a)}~A single column of $22$ force-sensing posts protrudes through a	smooth acrylic wall. \textbf{(b)}~A cornsnake (\textit{Pantherophis guttatus}) climbs by weaving through the posts; reflective markers on its back are tracked by an IR camera array. \textbf{(c)}~Each post couples to two load cells measuring vertical ($y$) and	horizontal ($x$) force. \textbf{(d)}~Three snapshots from a representative ascent: splined body (gray), markers (dark gray), and per-post forces (arrows); posts here are	$25$~mm long, spaced $50$~mm apart. \textbf{(e)}~Body-curvature kymograph (body position vs.\ time, colored by curvature $\kappa$); lateral bends propagate head to tail. \textbf{(f,g)}~Vertical \textbf{(f)} and horizontal \textbf{(g)} force on each post over time; summed vertical force tracks body weight and summed horizontal force stays near zero. \textbf{(h)}~Net torque about the center of mass is centered around zero.}
	\label{fig:schematic}
\end{figure*}

To study limbless climbing in a regime where the body cannot wrap around the substrate, we built a smooth acrylic vertical climbing wall with a single column of force-sensing posts\footnote{In the absence of posts, cornsnakes could not make forward progress on the acrylic surface and slid downward even at shallow inclines (SI~section~S1), confirming that the wall itself provided insufficient support. With posts present, however, all cornsnakes successfully ascended and descended the wall.}(see Methods). Figure~\ref{fig:schematic}a-c shows our apparatus, and  Fig.~\ref{fig:schematic}d shows a sequence of postures and associated force configurations during a typical ascent. Many simultaneous contacts were distributed along the body with forces generally pointed diagonally downward. The corresponding body-curvature kymograph (Fig.~\ref{fig:schematic}e) shows a head-to-tail propagation of lateral body bends that was typical for this wall configuration.  Horizontal and vertical forces were measured (in the plane of the wall) on each post through time (Fig.~\ref{fig:schematic}f,g). Using both time-resolved local kinematics and forces, we computed the net torque on the snake body about the center of mass through time (Fig.~\ref{fig:schematic}h). 

\begin{figure}[th!]
	\includegraphics[width=\columnwidth]{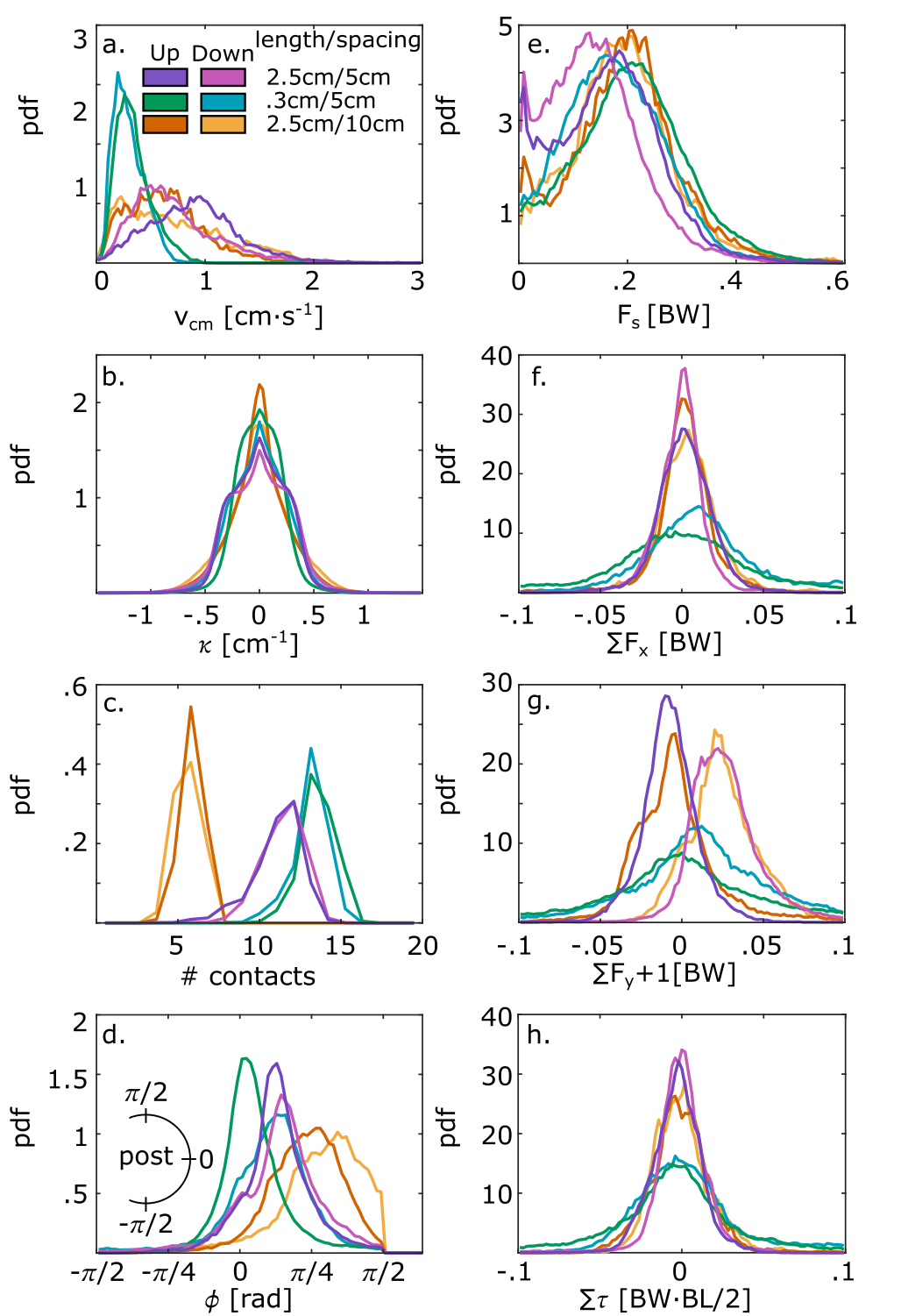}
	\caption{\textbf{Across substrates, snakes have variable kinematics but maintain force balance.} All panels are probability densities pooled across trials, for ascent and descent in three post configurations (legend, a). \textbf{(a)}~Center-of-mass speed: similar up and down, slower on shorter posts. \textbf{(b)}~Body curvature $\kappa$: similar across conditions. \textbf{(c)}~Number of simultaneous contacts: more on closely spaced posts, fewer when widely spaced. \textbf{(d)}~Radial contact location on each post, which shifts with condition. \textbf{(e)}~Total force per post: similar magnitude across conditions. \textbf{(f)}~Net horizontal force, net vertical force \textbf{(g)}, and torque about the center of mass \textbf{(h)} all center near balance; variability grows on shorter posts, where the force pattern is more erratic.}
	\label{fig:metrics}
\end{figure}

Cornsnakes were chosen as the focal species for this study as they are adept climbers, and their musculature and movements have been studied extensively in other contexts~\cite{astley_arboreal_2009,tingle_relative_2023,astley_effects_2007, jayne_why_2015}. 
Five cornsnakes were used in this study under three post configurations on the wall: $2.5$-cm posts spaced $5$~cm apart, $0.3$-cm posts spaced $5$~cm apart, and $2.5$-cm posts spaced $10$~cm apart. Each snake managed ascent and descent in all conditions (SI~section~S2) and Fig.~\ref{fig:metrics} summarizes their performance. 

For a given post configuration, speeds of ascent and descent were similar (Fig.~\ref{fig:metrics}a). On long posts, snakes used lateral undulation and climbed fastest (5--cm spacing Upward: SI Movie S1, 5--cm spacing Downward: SI Movie S2, 10--cm spacing Upward: SI Movie S3, 10--cm spacing Downward: SI Movie S4); for short posts, snakes used a mix of lateral undulation and concertina gaits (Upward: SI Movie S5, Downward: SI Movie S6), which resulted in slower speeds. This gait change is typical of snake behavior when surface features are reduced or absent~\cite{marvi_friction_2012,astley_arboreal_2009,jayne_why_2015}. Though several different complicated shapes were used in each trial and the body remained slightly straighter with short posts or wide spacing,  distributions of body curvatures were similar across conditions (Fig.~\ref{fig:metrics}b). Snakes formed many simultaneous contacts distributed along the body; however, the number of contacts varied with wall configuration: for closely spaced posts, they typically engaged $8$--$16$ contacts and contacted slightly more with shorter posts; with a wider spacing they had fewer contact points to work with and successfully climbed with $5$--$7$ contacts (Fig.~\ref{fig:metrics}c).

Radial contact location on the post varied across conditions. Snakes typically contacted the posts $30$--$45^\circ$ above horizontal, but during ascent, shortening the posts shifted contact locations toward the sides of the posts, whereas increasing post spacing shifted contacts toward the tops of the posts (Fig.~\ref{fig:metrics}d). Because force direction depended strongly on contact location, these shifts produced more lateral forces on shorter posts and more vertically aligned forces at wider spacing. Force magnitude per contact remained similar across conditions (Fig.~\ref{fig:metrics}e). However, due to an increased number of contacts, a greater total force was exerted on shorter posts (Fig.~\ref{fig:metrics}c) and was directed largely laterally, consistent with additional bracing on the more challenging substrate (SI~section~S3, Figure S2). By contrast, wider spacing reduced the number of contacts, requiring the remaining forces to align more vertically to maintain support. 

\textbf{Climbing is quasi-static.}
Despite marked differences in the kinematics and overall force pattern across conditions, the total force and torque remained nearly balanced at each moment in time for all conditions. Force directions, which depended on snake posture and contact location, alternated left and right along the body but the net horizontal force summed to zero overall (Fig.~\ref{fig:metrics}f). The measured vertical forces in each case were approximately equal to the snake's body weight (within a few percent, depending on the post configuration), indicating accelerations up and down the wall were small or absent (Fig.~\ref{fig:metrics}g). By calculating acceleration directly, we confirmed this was the case (SI~Figure~S3). We calculated the total torque about the snake's center of mass using a lever arm to each post and found this also remained small and centered on zero (Fig.~\ref{fig:metrics}h). Climbing therefore proceeded quasi-statically: at each instant the body approximately balanced net forces and torques. More formally, the snake satisfied three constraints:
\begin{equation}
	\label{eq:balance}
	\begin{aligned}
		F_{x,\mathrm{net}} &= \sum_i F_{x,i} \approx 0, \\
		F_{y,\mathrm{net}} &= \sum_i F_{y,i} - Mg \approx 0, \\
		\tau_{\mathrm{net}} &= \sum_i (\vec{r}_i - \vec{r}_{\mathrm{cm}}) \times \vec{F}_i \approx 0,
	\end{aligned}
\end{equation}

where $\tau_\mathrm{net}$ is the torque about the center of mass and $Mg$ is the animal's weight.
\footnote{Forces into/out-of the wall were measured in 17 experiments: they were small and not associated with displacements in that direction. We therefore conclude that the snake is below the static friction limit. See SI~section~S4.}.

The three constraints in Eq.~\ref{eq:balance} could in principle be met by as few as three contacts, yet the snakes maintained far more in nearly every condition (Fig.~\ref{fig:metrics}c). In this scenario, the snake can access a wide range of force distributions that all satisfy the quasi-static constraints. Even in the short-post condition, where transient departures from balance were more common, the net force and torque distributions remained centered within a few percent of balance, indicating that most of the time, the body nearly maintained quasi-static equilibrium. However, global force and torque balance do not uniquely determine how individual contacts contribute to locomotion. In particular, they do not reveal whether contacts act to oppose motion, as in passive friction, or whether they contribute to forward progression. Therefore, to determine the minimal requirements that are mechanically necessary for limbless climbing, we built two complementary models of climbing on a post array.

\begin{figure*}[th!]
	\includegraphics[width=0.9\textwidth]{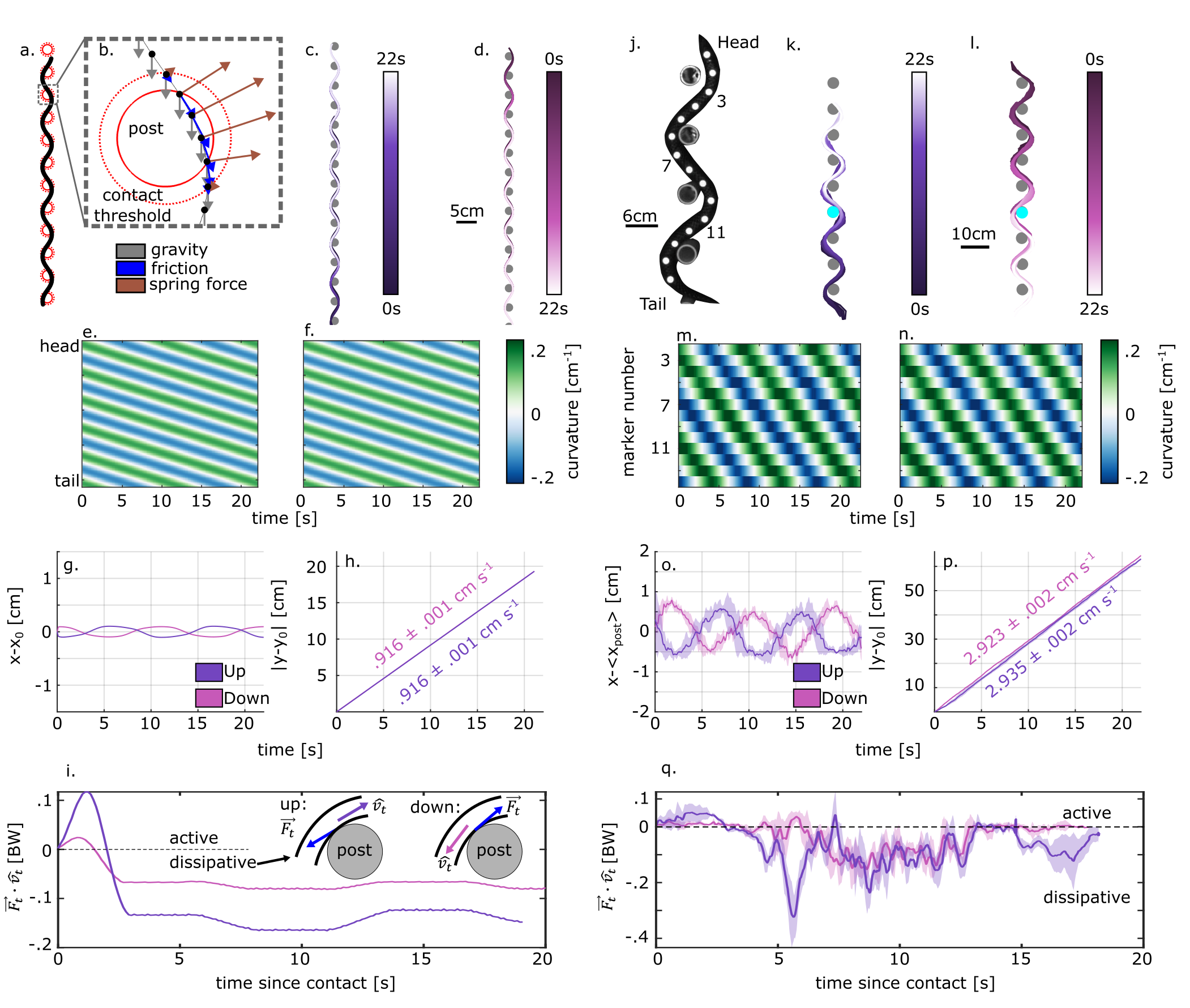}
	\caption{\textbf{Computational and robotic models reveal the minimal requirements for climbing.} \emph{(a--i) Computational model.} \textbf{(a)}~A chain of segments propagates a prescribed serpenoid wave through the post array; \textbf{(b)}~each contacted segment feels gravity (gray), Coulomb friction (blue), and a repulsive spring force (brown). The model climbs both up \textbf{(c)} and down \textbf{(d)}, cycling through serpenoid shapes as bends propagate head to tail identically in both cases. (curvature kymographs: \textbf{e}, up; \textbf{f}, down). \textbf{(g)}~Lateral and \textbf{(h)}~vertical center-of-mass displacement show negligible net horizontal motion over	a cycle and constant-velocity vertical advance---consistent with quasi-static climbing. \textbf{(i)} The tangential force at a single contact opposite the tangential velocity showing dissipation after a brief adjustment period. Note: here all contacts behave the same. Inset: for dissipative contacts, tangential force opposes local velocity. \emph{(j--q) Open-loop-controlled robot.} \textbf{(j)}~A chain of servo motors propagates a serpenoid wave, with reflective markers tracked at each joint. The robot climbs up	\textbf{(k)} and down \textbf{(l)} using only 3--4 contacts, far fewer than the snake. Cyan dots in \textbf{(k,l)} represent the location of the force-sensitive post. Curvature kymographs (\textbf{m}, up; \textbf{n}, down) and center-of-mass displacement (\textbf{o}, lateral; \textbf{p}, vertical) again indicate constant-velocity, quasi-static climbing. \textbf{(q)} The average tangential force against tangential velocity at the force-sensitive contact again showing dissipation after a brief adjustment.}
	\label{fig:models}
\end{figure*}

\textbf{Computational and robotic models establish a minimal climbing baseline.}
To understand how snakes utilize surface feature contacts, we first constructed a quasi-static computational model of a climbing snake interacting with posts in its environment. The the snake body is represented as a chain of connected segments undergoing a prescribed serpenoid wave (see Methods). As the body moves through the array of posts, each segment experiences up to three environmental forces. At posts, an effective spring force normal to the body enforces local contact geometry and kinetic Coulomb friction opposes local motion across the contact, while gravity acts on all segments (Fig.~\ref{fig:models}a,b). At each instant, the whole-body translation was determined by enforcing net force and torque balance (see Methods). Surprisingly, this simplified model produces successful climbing behavior both up and down establishing an important minimal baseline: passive contact forces are sufficient to produce quasi-static climbing.

The success of passive contact dynamics in our computational model inspired us to design a robotic model to confirm the viability of this scheme (Fig.~\ref{fig:models}j). Because climbing is quasi-static, balance requires satisfying only the three constraints listed in Eq.~\ref{eq:balance}, which provides a theoretical minimum number of contacts needed
By propagating a serpenoid wave down the body, the robot was able to climb up and down a vertical post array with no environmental sensing or mechanical force feedback (Fig.~\ref{fig:models}k,l) (Upward: SI Movie S7, Downward: SI Movie S8). The robot maintained only $3$--$4$ contacts at a time, many fewer than the snake, confirming a near-minimal amount of contacts required for successful behavior. The curvature waves were identical in the two directions (Fig.~\ref{fig:models}m,n), so climbing direction required no change in gait (see Methods). The robot's kinematics were also consistent with a quasi-static system as the horizontal center of mass was centered around zero (Fig.~\ref{fig:models}o) and the vertical center of mass advanced at nearly constant velocity (Fig.~\ref{fig:models}p), indicating little to no acceleration. Though our force sensors were not able to support the weight of the robot, we designed a wall where one post was equipped with a 6-axis force/torque sensor to measure forces exerted by the robot throughout each climb. To ensure that we captured force dynamics throughout a climb, we mounted this sensor to a middle post (fourth from the bottom) and measured the forces before, during, and after the robot interacted with that post. After a brief adjustment period following initial contact, interactions with this post were generally dissipative, confirming that climbing is physically allowed with friction dominating at contacts (Fig.~\ref{fig:models}q).

The simulation and robot are complementary: the computational model, with friction-only contacts, shows that a propagating body wave and passive friction are sufficient in principle, while the robot confirms this physically and with only a few contacts. Together they establish a minimal baseline---open-loop body-wave propagation with passive, frictional contacts is enough to climb quasi-statically, requiring neither the snake's redundant contacts nor feedback.


\textbf{Snakes actively redirect forces during ascents.}

To test whether snakes use this minimal strategy, we calculated the energy transfer direction at each contact. Sliding contacts were considered \emph{dynamic} where the local velocity of the body exceeded $3$~mm/s. If the local velocity did not meet this threshold, the contact was considered \emph{static}. For each dynamic contact, we decomposed the measured force and local velocity into components tangential and normal to the average radial contact location $\phi$. Given the extended nature of the snake body, we used a distance-weighted average of angular contact locations to define a contact centroid; a similar approach was used to estimate contact sliding velocity (SI~section~S5). 
We define the tangential contact power to be $P_t = F_t\cdot v_t$, where $F_t$ and $v_t$ are the force and velocity components tangential to contact location defined by $\phi$.
$P_t$ is negative when the tangential force opposes local sliding direction, indicating a \emph{dissipative} frictional interaction; $P_t$ is positive when tangential force aligns with the sliding direction---and does positive work on the body---which passive Coulomb friction cannot.

\begin{figure}
    \captionsetup{width=\columnwidth}
    \includegraphics[width=\columnwidth]{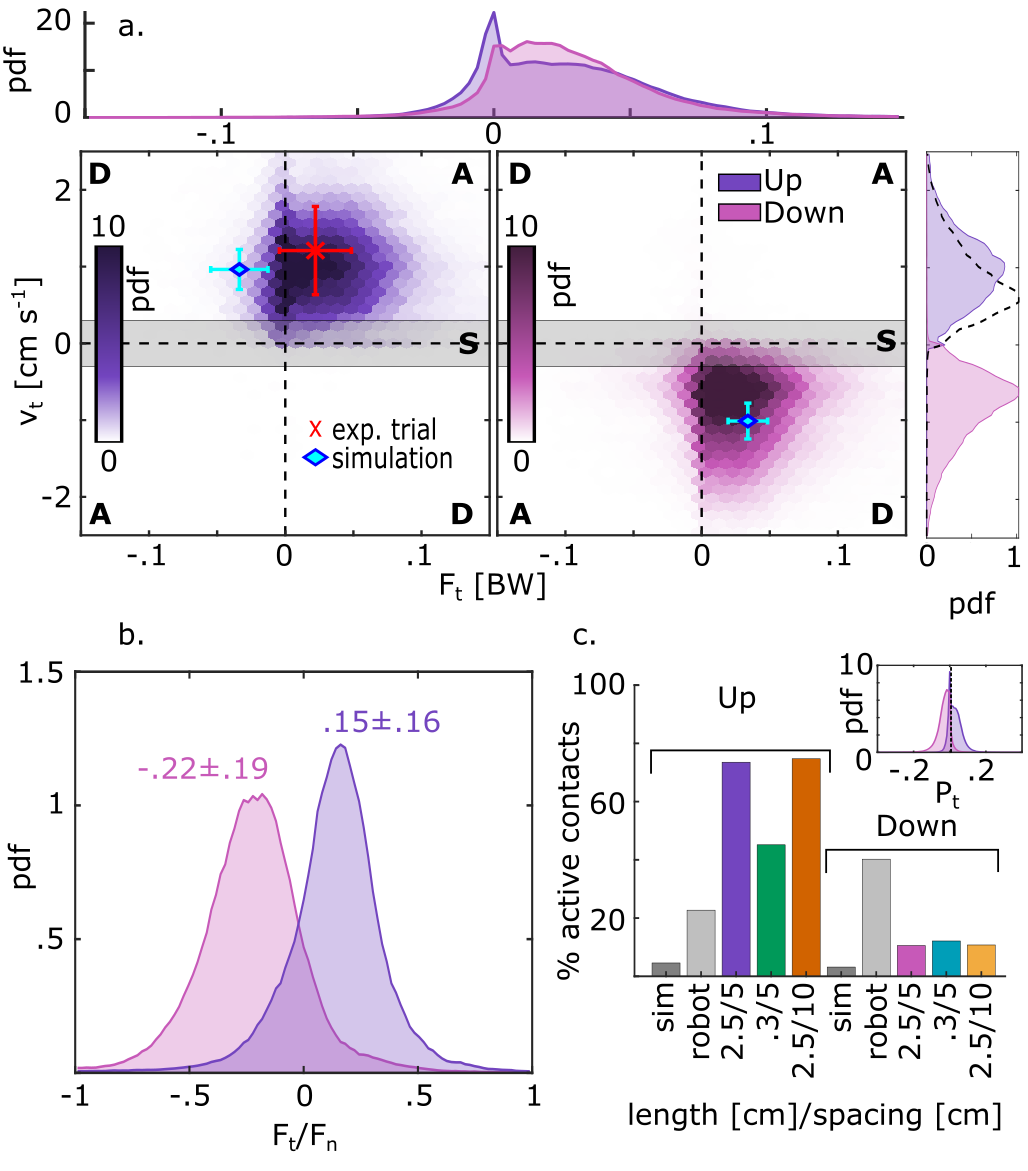}
    \caption{\textbf{Downward climbs have mostly dissipative interactions at contacts while upward climbs are active.} \textbf{(a)} Heatmaps of $F_t$ and $v_t$ with quadrants marked ``A'' and ``D'' respectively indicate active and dissipative contacts respectfully. Dissipation in downward climbs matches simulated predictions while upward climbs, including the trial that the model was based on (red ``x") involve active contacts. Distributions of tangential forces and velocities are shown atop and to the right of the main plots, respectively.\textbf{(b)} The ratio of tangential and normal forces in downward climbs matches closely with the $\mu$ predicted from dissipative contacts. In contrast, this ratio is nearly opposite in upward climbs. \textbf{(c)} The percentage of active contacts in each condition is much higher for upward climbs while downward climbs lie closer to model predictions.}
    \label{fig:power}
\end{figure}

At most contacts during experimental descents, the snake's tangential velocity sliding down the wall was opposed by a tangential force on the snake directed up the wall ($P_t < 0$; Fig.~\ref{fig:power}a), consistent with the dissipative interactions predicted by the simulation and robot. Assuming these dissipative contacts are purely frictional gives an effective kinetic friction coefficient $\mu \approx 0.22$. Upward climbs, however, exhibited reversed dynamics: at most contacts, the snake generated a tangential force differing from the passive-friction prediction by roughly $2\mu F_n$, overcoming the resisting friction and driving an equal force in the opposite direction to do positive work on the body.

We refer to these positive-power contacts as \emph{active}; meaning that the measured tangential force does positive mechanical work on the body relative to local sliding ($P_t > 0$). Critically, $P_t > 0$ is not a generic consequence of propagating a body wave through the array. Our open-loop-controlled robot, which does exactly that with no sensing or control, produces dissipative contacts on average (SI~Figure~S7). Active contacts therefore reflect an animal-specific contribution beyond this minimal feed-forward mechanism. We stress that ``active'' remains an energetic statement, not a claim about neural control: our measurements cannot distinguish whether the animal redirects force through sensory feedback, a feed-forward muscular program, or the passive mechanics of a deformable body cross-section.
More than $40\%$ of dynamic contacts were active during ascent in every environmental condition, whereas active
contacts were much less common during descent, which more closely matched simulation predictions (Fig.~\ref{fig:power}c). Shortening the posts modestly reduced the prevalence of active contacts, consistent with the shift toward a more concertina-like gait with a greater fraction of static contacts. The robot-post interaction was also occasionally active, but only briefly, when stick--slip dynamics deflected the contact force in the direction of local motion.

The energetic contrast between ascent and descent arose primarily from the reversal of sliding direction rather than from a change in tangential force. Tangential-force distributions were similar for both climbing directions, whereas the tangential-velocity distributions nearly coincided after a sign reversal (Fig.~\ref{fig:power}a, dashed black line). Consequently, similarly directed forces tended to oppose sliding during descent but align with sliding during ascent, shifting contact power from predominantly negative to frequently positive. Because both the simulation and robot could climb upward using mainly dissipative contacts, this positive work was not required for ascent itself. Instead, it points to an additional force-redistribution strategy used by the animals, raising two questions: how do snakes generate these active interactions, and what functional advantage might they provide?

\textbf{New contacts trigger stereotyped, balance-preserving force redistribution.}

\begin{figure*}[t!]
	\includegraphics[width=\textwidth]{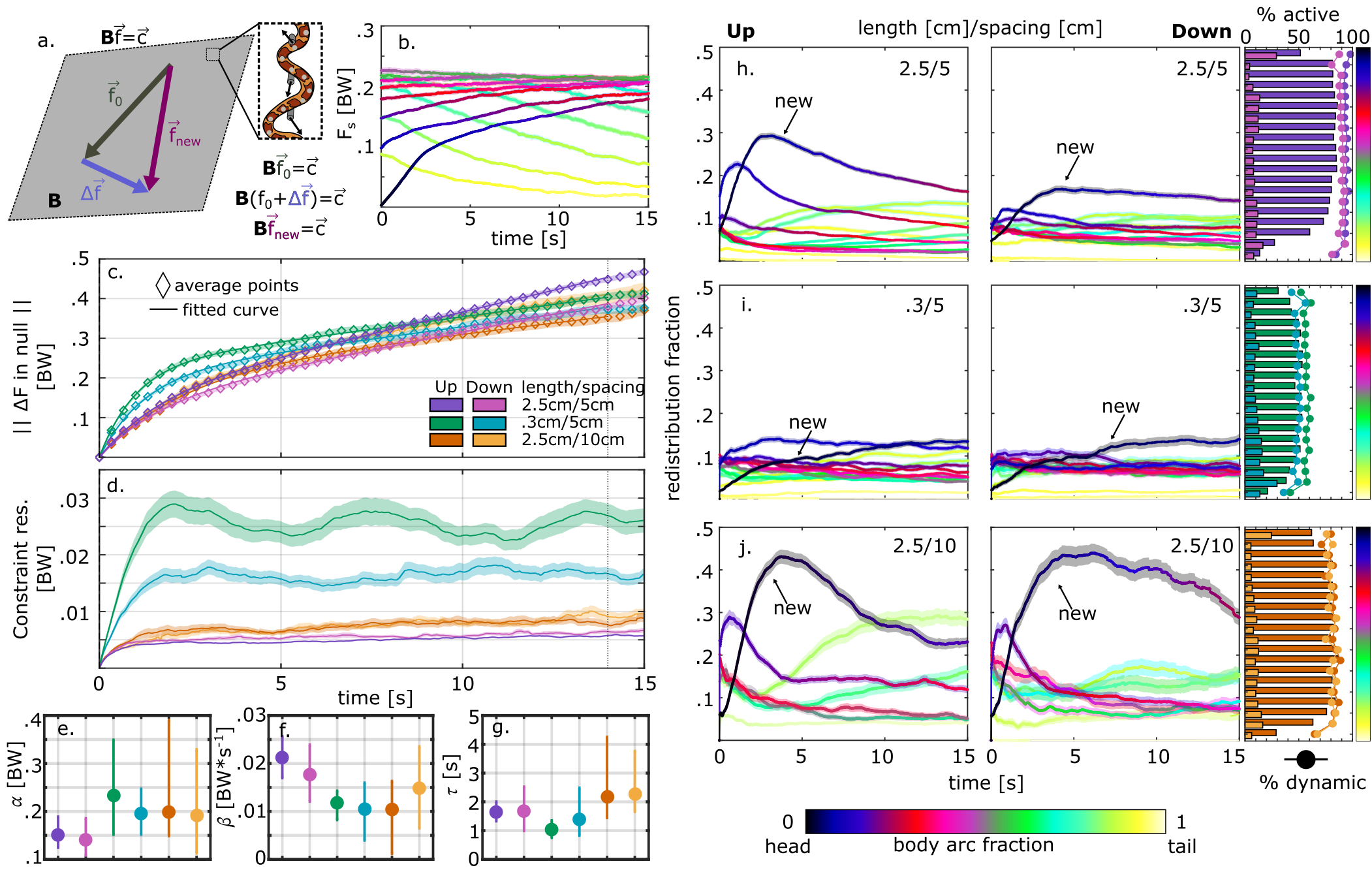}
	\caption{\textbf{Force redistribution is stereotyped and balance-preserving.} \textbf{(a)}~The balance constraints $\mathbf{B}\vec{f}=\vec{c}$ define a space of balanced force configurations; a rearrangement $\Delta\vec{f}$ in the nullspace of $\mathbf{B}$ moves between them ($\vec{f_0}\!\to\!\vec{f}_{\mathrm{new}}$) without disturbing balance. \textbf{(b)}~Contact-triggered average of the force on each post: force
		ramps to a stable value on the new contact and its near neighbors, a stereotyped incorporation. \textbf{(c)}~Magnitude of the balance-preserving redistribution $\lVert\Delta\vec{f}_{\mathrm{null}}\rVert$ over time (average points and fitted	curves), all six conditions. \textbf{(d)}~Residual force outside the nullspace (the balance violation): it rises and then holds at a small, steady level---largest on short	posts, but within a few percent of body weight. \textbf{(e--g)}~Fitted parameters across	conditions: amplitude $\alpha$ \textbf{(e)}, drift rate $\beta$ \textbf{(f)}, and timescale $\tau$
		\textbf{(g)}. (h--j)~Fractional contribution of each contact to the redistribution in (c), colored by body position (head to tail), climbing up and down; the new contact is labeled. The redistribution concentrates near the new contact on long, closely spaced posts, spreads along the body on short posts, and is dominated by the new contact at wide spacing. The plots on the right track the percentage of contacts along the body that are dynamic and active showing how local force rearrangements affect energetics.}
	\label{fig:null}
\end{figure*}

The contact-power analysis describes the mechanical role of each contact in isolation, but climbing stability depends on how the snake coordinates several contact interactions at once. With many redundant contacts, several different force configurations $\vec{f_0}$ satisfy force and torque balance and make up a ``balance space" $\textbf{B}$ (Fig.~\ref{fig:null}a). This wide range of solutions allows the animal to change the load on any individual post without necessarily departing from force balance. We therefore asked how snakes redistribute forces across the contact network each time a new post is engaged.

Incorporating a new contact is not an independent action; it triggers a repeatable redistribution of force through the contact network as force ramps up on the new post and eventually reaches a stable, stereotyped value (Fig.~\ref{fig:null}b). We represent the set of contact forces as a vector $\vec{f}$ and the quasi-static balance conditions as $\mathbf{B}\vec{f} = \vec{c}$, where the three constraints fix the net horizontal force, the net vertical force, and the net torque about the center of mass. Any rearrangement, $\Delta \vec{f}$, to this force configuration separates into two mechanically distinct parts: a balance-preserving part lying in the nullspace of $\mathbf{B}$, leaving the constraints $\vec{c}$ satisfied, and a residual part that violates whole-body balance. The nullspace is precisely the freedom that redundancy provides: within it, the animal can reshape how force is shared among posts without perturbing its equilibrium. We quantify rearrangements across trials by defining $t=0$ to coincide with the onset of each new contact and tracked the change in the contact-force vector relative to its value at that instant, focusing first on the balance-preserving (nullspace) component, $\Delta\vec{f}_{\mathrm{null}}(t)$ (see Methods, SI~section~S6).

Across all conditions, the magnitude $\lVert\Delta\vec{f}_{\mathrm{null}}(t)\rVert$ rose rapidly after a contact formed and then continued to drift as the animal advanced through the array (Fig.~\ref{fig:null}c). We summarized this response with a saturating exponential plus a linear ramp,
\begin{gather}
	\lVert\Delta\vec{f}_{\mathrm{null}}(t)\rVert
	= \alpha\left(1 - e^{-t/\tau}\right) + \beta t ,
	\label{eq:redist}
\end{gather}
where $\alpha$ is the amplitude of the rapid, balance-preserving redistribution recruited by the new contact, $\tau$ is the characteristic time over which that transient settles, and $\beta$ is the rate of the slower, ongoing reassignment of force as body position, contact geometry, and moment arms change during the climb. We interpret this fit as a description rather than a mechanistic claim: the exponential term captures the transient response to a newly available contact, and the linear term reflects the animal's continued motion through a changing mechanical environment. 

The part of the rearrangement that violated force and torque balance rose over the first second or two and then held at a small, steady level rather than decaying away (Fig.~\ref{fig:null}d). This also provides a proxy for how closely the system remains quasi-static. With longer posts, less than $1\%$ of the force was outside of balance, however, shorter posts led to more variable force patterns and greater departures from being quasi-static. 

Across the six conditions, the transient amplitude $\alpha$ was smallest on long, closely spaced posts and larger when posts were shortened or spaced farther apart (Fig.~\ref{fig:null}e), though these differences were comparable to the fit uncertainties. The ongoing drift rate $\beta$ showed an opposite, and clearer, pattern: largest on long, closely spaced posts and roughly half as large on shorter or sparser ones (Fig.~\ref{fig:null}f), though uncertainties overlapped here as well. The rearrangement timescale $\tau$ remained similar in each condition ($\sim1$-$3$~s) (Fig.~\ref{fig:null}g). Together, these results point to a trade-off. With dense, reliable contacts, each new contact triggers a small force rearrangement (smaller $\alpha$) but forces must propagate down the body at a higher rate (larger $\beta$). When the surface features are shorter or sparser, the opposite is required: each new contact requires a large force rearrangement (larger $\alpha$) but loads can be transferred between body segments slower (smaller $\beta$).

\textbf{Body segments are recruited for rearrangement based on the environment.}

Having established that new contacts trigger mostly balance-preserving redistributions, we next asked which parts of the body contribute most to these force changes. Although redistribution occurs across the contact network, each force is generated at a localized body--post interaction. The spatial pattern of contributions therefore reveals whether contact incorporation is concentrated near the newly formed contact or distributed more broadly along the body. 

To investigate, we calculated the fractional redistribution contributed by each contact. On long, closely spaced posts, the established contact just behind the new one initially carried the most rearrangement, but later on the new contact became the dominant contributor to the rearrangement. Middle and hind segment contributions were small and steady until the tail released the trailing post (Fig.~\ref{fig:null}h). This concentration was strongest during ascent and was responsible for redirecting the tangential force; the region where more dynamic contacts were active occurred just behind the head-most region of the body. 

For descent the same trend held, but contributions were more evenly shared along the body while the redistribution maintained dissipative interactions. When posts were shortened, no body segments emerged as relatively strong redistributors (Fig.~\ref{fig:null}i). Instead, when using concertina-like gaits with multiple sliding and static contacts on the body simultaneously, more body segments were actively involved and contributed equally to the force redistribution. We observed an increase in mid-body adjustments in this environment: on these substrates, the mid-body sometimes flipped over a post, engaged it from the opposite side (Movie~S9), and recruited body regions that were previously uninvolved. When posts were spaced farther apart, the new contact made a larger and more persistent contribution, as expected when each contact carries greater weight in a sparser network (Fig.~\ref{fig:null}j).

Together, these results show that each new contact is incorporated through a redistribution of force across the existing network that is largely balance-preserving, but whose size and spatial spread depend on the density and reliability of the available surface features. The limbless body thus uses redundancy not only as excess support, but as a mechanical space in which local forces can be reorganized while maintaining stability.

\section*{Conclusions}

For an elongated, limbless body, generating stabilizing and propulsive force on a flat, vertical surface requires much more than generating friction at isolated footholds. Instead, it is important to manage a redundant contact network as a whole to distribute force across contacts as new ones are established. Our computational and robotic models proved that passive environmental contact forces coupled with simple body-wave propagation is sufficient for climbing, but cornsnakes operate in a much richer regime. They engage far more contact points than balance strictly requires, actively redirect force beyond passive friction during ascent, and smoothly incorporate each new contact through a highly coordinated, balance-preserving redistribution.

This gap between what is mechanically sufficient and what the animals do reveals both an opportunity and a demand of the limbless body plan. A continuous body can engage many features at once, opening a redundant space of force distributions where coordinated signals can utilize many different solutions to maintain balance~\cite{todorov_optimal_2002}. But this redundancy is not automatically useful: an elongate body does not by itself guarantee climbing. Kingsnakes failed on the same wall (Movie~S10). Cornsnakes succeeded by organizing their scattered contacts into a balanced, force-sharing network. With active control localized on segments of the body at each point in time, the exact mechanism by which force is redirected remains unknown. Finely controlled muscles connecting the ribs and backbone to the skin are also highly localized and we posit that these may be responsible for redirecting force~\cite{Gasc1981}. Whether force redirection is neurally controlled or a passive consequence of body deformations falls outside the scope of our data. Whatever the mechanism, doing positive work at a contact rather than merely resisting it requires actuation coordinated with the local sliding velocity, implying finely distributed force sensing along the body~\cite{tytell_spikes_2011}.

Efficient motor control involves conserving energy by ignoring redundant dimensions~\cite{todorov_optimal_2002}. In extreme environments, however, energy efficiency may be secondary to robust locomotor performance and stability. Thus, climbing snakes may utilize a system where redundant contacts provide a wide space of solutions to prevent perturbations that may result in falls. The redundancy itself is a principle that reaches well beyond snakes. Whenever a task can be met in more ways than it constrains, e.g., reaching for an object with a multi-jointed arm~\cite{bernstein_coordination_1967}, grasping with many fingers~\cite{bicchi_hands_2000}, or stabilizing the body over uneven terrain~\cite{daley_running_2006}, the surplus degrees of freedom span a subspace where activity can vary without affecting the task, which motor systems exploit to absorb perturbations while protecting task-relevant variables~\cite{scholz_uncontrolled_1999}. Here, the snake reveals an example of redundancy: a space of internal forces that a multi-point grasp can apply without changing the net wrench on the object~\cite{murray_mathematical_1994}. This allows versatility in internal force patterns which they use to hold a climbing body balanced as contacts come and go. That snakes maintain such a network at the cost of many
extra contacts reflects a trade-off familiar across biology and engineering---between the economy of a minimal solution and the fault tolerance of a redundant, over-constrained one~\cite{kitano_biological_2004,carlson_complexity_2002}. Where a single slipped contact can mean a fall, holding whole-body balance to a small, bounded error may be worth the cost, consistent with the tendency of biological systems in high-stakes environments to favor robust redundancy over minimal viability~\cite{alexander_factors_1981}.

These results lend themselves to further development of robotic templates which coordinate signals across a network to produce movement~\cite{ijspeert_central_2008,aguilar_review_2016,ramdya_neuromechanics_2023,biswas_mode_2023}. Here we have demonstrated the minimum required controls given that the geometry of the propagated wave matches reasonably with the post spacing. Snake-like climbing gaits have been replicated previously allowing these robots to traverse inclined poles~\cite{tang_arboreal_2017, rollinson_pipe_2016}. However, our results work toward expanding the repertoire of passible terrain for these devices.

These principles place the repeated evolution of arboreal climbing in snakes within a broader picture of contact-rich locomotion~\cite{wiens_why_2006,jayne_muscular_1988}. Low-curvature surfaces cannot be encircled like branches, yet they offer scattered features that a body can recruit. Limbless animals find success in these situations by aggregating features into a malleable, balance-preserving contact network which may be key to managing habitats containing them. By separating the mechanics that make climbing possible from the strategies animals actually use, this work points to the redistribution of force across a redundant contact network---rather than friction at isolated footholds---as a central feature of limbless climbing.

\section*{Materials and Methods}

\textbf{Study subjects.}
Five juvenile cornsnakes were used in this study. All individuals were between $16$ and $24$ months old, weighed between $77$ and $114$~g, and were between $69$ and $82$~cm long (SI~section~S8). The snakes were fed every $7$ days. Individuals were eligible to be used for trials on a given day if they: had not eaten in the last $24$ hours, were not in shed, and had not been used the previous day. Each individual was not allowed to complete more than $10$ trials per day and, in total, each completed between $38$ and $54$ trials over the whole study amounting to $227$ trials in total (SI~section~S2). All experiments took place in the same room that the snakes were housed. The room temperature was consistently $75$-$80^\circ$F and the ambient relative humidity fluctuated from $15$-$30\%$. All experiments were performed in accordance with Emory University IACUC protocol 202100179.

\textbf{Experimental apparatus.}
The climbing apparatus used for the experiments consisted of a flat, smooth piece of acrylic, $152$~cm ($5$~ft.) tall by $30$~cm ($1$~ft.) wide. The wall was mounted vertically facing our array of marker tracking cameras. Protruding from the wall was a single column of metal posts, $6$~mm in diameter. The posts were covered with a smooth, 3D-printed sleeve $8$~mm in diameter. Foam padding was placed at the base of the wall to ensure a soft and safe landing in the event of a fall. Posts were mounted behind the wall and protruded through holes in the acrylic $10$~mm in diameter so that the posts did not touch the inside of the holes. Each post was mounted to custom metal brackets which accommodated two TAL221 load cells from Sparkfun with a $500$~g load capacity. One load cell was oriented to measure vertical forces on the post while the other was oriented to measure horizontal forces. Each load cell was connected to its own HX711 analog-to-digital signal amplifier which was connected to an Arduino Mega for data collection. There were $22$ posts on the wall which required $44$ load cells, $44$ amplifiers, and $2$ Arduino Megas to collect all of the force signals. We were able to collect vertical and horizontal force measurements from all posts simultaneously at $\sim10$~Hz.

We were able to manually move the acrylic wall back and forth to change the length of the posts protruded. In these experiments, posts were mounted $50$~mm apart and were set to lengths of $25$~mm and $3$~mm. The length of the 3D-printed sleeves on each of the posts matched the length of post protruding from the wall. 

To record 3D kinematic data, we used an array of 6 Optitrack Primex22 cameras recording at 60 Hz, 5 tracking reflective markers on the back of the snake and 1 recording video. Kinematic data and force data were later synchronized (see Data Analyses below) in MATLAB for analysis. 

\textbf{Force-sensor calibration.}
Each load cell is designed with strain gauges to output a signal in $\mathrm{mV}$ linearly proportional to the loaded weight on the post (measured initially in grams). Each load cell was calibrated independently by hanging a series of known masses from each post and calculating the slope of the output signal. Horizontal load cells were loaded horizontally with the help of a pulley system. Because each pair of load cells were coupled, we found cross-signals between the two, i.e., a purely vertical load would affect the horizontal load cell's reading. To account for this, vertical and horizontal forces were each calibrated using both load cells and both the linear and quadratic responses to the series of static loads on each (SI~section~S9). 

\textbf{Experimental procedure.}
Each day, 2--3 snakes were made available for experiments to perform upward and downward climbs given that they met the criteria mentioned above. Before performing the experiments, the snakes were marked with $30$--$40$ markers $6$~mm ($1/4$~in) in diameter cut from silver reflective tape and adhered at roughly even intervals down each snake's back. A trial order was randomly generated dictating the order to use each individual and climb direction. 

Before performing trials on a given day, the ends of the posts and four corners of the wall were marked with the same markers as the snakes and recorded with the cameras for $3$~s. From this we were able to show the snake position relative to the post positions and the wall position. Markers on the wall and posts were removed before experiments took place. The wall and posts were kept in the same position throughout the day.

A trial started when the individual snake is introduced to the wall (at the bottom of the post column for upward climbs, at the top for downward). Once the snake climbed fully up or down the column of posts, the trial was completed. If the desired behavior was not observed, the snake was repositioned for another attempt. If the snake remained stationary or paused during climbing, we tapped their tail to elicit an escape response. If after $10$ minutes no climbing behavior was observed, the trial was skipped and the next random trial in the order was performed. Snakes rested while other individuals performed trials. If a snake was scheduled for trials back-to-back, no rest was allowed if the trial lasted less than 2 minutes. For longer trials, the individual was allowed to rest for at least $10$ minutes in between trials. Individuals were subjected to no more than $10$ trials and/or $10$-minute periods with no desired behavior per day.

\textbf{Quasi-static climbing model.}
To ask what is minimally required for a limbless body to climb through a sparse set of contacts, we built a simple model of a snake weaving between posts [LINK TO CODE]. We represented the body as a chain of $N = 220$ rigid segments of equal mass $m = M/N$, where $M$ is the total body mass. Each segment had a length $\ell_s = L/(N-1)$, and the total body length was $L$. Rather than solving for the body shape, we prescribed it as a traveling serpenoid wave---a posture in which the curvature, $\kappa(s,t)$, varies sinusoidally along the body and travels from head to tail at constant speed, $v_c$. The curvature is given by
$$\kappa(s,t) = \kappa_m \sin\left[\frac{2\pi}{\lambda_s}(s - v_c t)\right],$$
where $s$ is arc-length position along the body, $t$ is time, $\kappa_m$ is the peak curvature, $\lambda_s = L/n_w$ is the wavelength, given by the body length divided by the number of waves $n_w$. Integrating the curvature along the body gives the segment tangent angles in the laboratory frame,
$$\theta(s,t) = \theta_0 - \frac{\kappa_m \lambda_s}{2\pi}\cos\left[\frac{2\pi}{\lambda_s}(s - v_c t)\right],$$
where $\theta_0$ orients the body along the direction of climbing. We chose the waveform parameters to approximate the body shapes we observed during climbing, and we set $v_c$ so that the resulting center-of-mass speed matched our measurements.

Every segment experienced gravity, $\vec{F}_g = m\vec{g}$. A segment experienced additional forces only when it overlapped with a post. Specifically, when the penetration depth $\delta$ (the distance by which the segment intrudes past the post's effective radius) exceeded zero, the segment experienced a contact reaction force and friction; a segment that was not in contact ($\delta = 0$) experienced only gravity. We resolved these contact forces in the body frame, using the unit vectors normal ($\hat{n}$) and tangent ($\hat{t}$) to the segment. The reaction force was a repulsive, one-sided penalty force that resisted interpenetration, given by
$$\vec{F}_N = K\delta\hat{n} \qquad (\delta > 0).$$
This force grew linearly with the overlap $\delta$ (an exponent of one is appropriate for cylindrical contact), with stiffness $K$, and pointed along the segment normal so as to push the body out of the post. Kinetic Coulomb friction acted along the body and opposed sliding,
$$\vec{F}_f = -s\mu_k |\vec{F}_N|\hat{t} \qquad (\delta > 0),$$
where $\mu_k$ is the coefficient of kinetic friction, the normal load is the reaction force magnitude $|\vec{F}_N| = K\delta$, and $s = \mathrm{sgn}(v_\parallel)$ is the direction in which the segment slides along its own axis. In other words, the perpendicular direction captured the contact reaction force and the parallel direction captured friction.  Gravity was the only force with components along both the parallel and perpendicular directions.

A rigid post can react only along its own surface normal, which coincides with the segment normal $\hat{n}$ when the body lies tangent to the post, as it does for the contacts we observed. We deliberately apply the reaction along $\hat{n}$ rather than along the exact post normal: directing it along the post normal would couple the force direction to the center-of-mass displacement we solve for at each step, which generally prevents the body from reaching exact force balance for a prescribed shape, whereas the body-normal form leaves the force directions fixed by the posture and admits an exact quasi-static balance.

Consistent with our observation that climbing snakes accelerate very little (Fig.~\ref{fig:metrics}), we treated the body as quasi-static, requiring the net force and net torque to vanish at every instant,
$$\sum_i \vec{F}_i = 0 \qquad \text{and} \qquad \sum_i (\vec{r}_i - \vec{r}_{cm})\times\vec{F}_i = 0,$$
where $\vec{F}_i$ is the total force on segment $i$ and $\vec{r}_i - \vec{r}_{cm}$ is its position relative to the center of mass. At each time step we held the prescribed shape fixed and solved for the small rigid-body displacement of the center of mass---two translations and one rotation---that brings the net force and torque closest to zero, normalizing the torque by the body length $L$ so that it scales comparably with the forces. The waveform parameters reported in the main text are those that best reproduced the body shapes we observed.

\textbf{Robotic model.}
To test the viability of passive contact mechanics and near-minimal contacts in a physical system, we designed an open-loop-controlled robotic model. The robot consists of $15$ servo motors linked together with 3D-printed brackets to create a motor chain with a joint between each motor pair. We sinusoidally varied the joint angles to generate a serpenoid wave. With $2$ waves on the body and joint angles between $135^\circ$ and $180^\circ$, the robot was able to climb up and down a vertical post array with posts made out of $41$-mm-diameter PVC pipes spaced $127$~mm apart. One post in the center of the array was connected to an ATI Mini-40 6-axis force/torque load cell to measure forces at a single contact throughout the entire robot-post interaction. Waveform parameters were chosen so that the robot had only $3$--$4$ contacts throughout each trial.

\textbf{Data analyses.} 
All data were analyzed using MATLAB. Data and code can be found here: \url{https://osf.io/x5rdu/overview?view_only=3a8c024555c747d0a15d5f2bab60ec9e}. Force data were synchronized to points in the marker tracking data using timestamps on each force point and a timestamp for the start of the marker tracking data. Because the force data was recorded at a lower frequency than the cameras, the data points were linearly interpolated to match the $60$~Hz kinematic data. Force data were calibrated using calibration matrices for each individual load cell and normalized by the weight of the particular snake. 


During climbing assays, the cameras recorded the 3D positions of the $30$--$40$ markers along the back of the snake. Most markers were tracked for the entire duration of each trial, but occasionally some were dropped and reappeared. These markers were re-linked by finding two markers that disappear/reappear closest to the same spatial location. Any trials where markers were dropped and unable to be re-linked were truncated to the longest time span where all markers were present. If the longest time span was greater than $20$~s, the trial was excluded from analyses. Once linked through time, markers were ordered by proximity along the body, starting with the one at the greatest height. 

We used the position of the wall to transform all of our spatial data into a consistent coordinate system. The markers recorded at corners of the wall were used to fit a plane to the wall. By building an orthonormal basis matrix to describe the plane, we  linearly transformed all of the snake, post, and wall marker points relative to this plane. As a result, the plane of the wall became the $xy$ plane where $y$ is directed vertically and $x$ is directed horizontally. The $z$-direction describes how far away from the wall the markers are (SI Figure S5). We also smoothed the data using a 1D Gaussian filter with a standard deviation $\sigma=20$~time points and a radius of $4\cdot\sigma$ over time to eliminate small perturbations below the noise threshold of the camera array ($\sim 0.2$~mm).

We calculated the position of the body segments relative to the posts and examine the body segments used to apply the force detected by each post and their shape at each point in time. We used Modified Akima Interpolation to obtain a set of $200$ cubic spline points along the snake's backbone at each point in time. The curvature of the body was calculated from these spline points through time using the Pratt method to fit circles to local regions of 11 points. From here, we used a distance threshold of $17$~mm to identify which body segments were close to which posts. We also used a force threshold of $0.21$~g, which we found to be $3$x the standard deviation of the sensor readings at rest, to show when a contact was established.

We chose to focus on segments of trials in which snakes were actively making upward or downward progress during climbs. Thus, if the snake stopped, i.e.,  $v_{\mathrm{cm}} < 1$~mm/s for at least $2$~s,  we truncated the trial to the longest time span with no pauses. Lengths of $3$~s either before or after a pause were also eliminated to avoid sudden speed ups and decelerations. If the longest time span with no pauses was not at least $20$~s, the trial was excluded. 

Kinematic quantities at each contact such as the radial contact position $\phi$ and tangential velocity $v_{t}$, were calculated using a weighted average of the quantity measured at every body segment within the distance threshold of the contact. Weights were assigned based on the distance from each point to the contact threshold (SI~section~S5). 

A balance space was constructed based on possible arrangements of applied forces that satisfied the constraints of force and torque balance. We defined a nullspace of this balance consisting of force rearrangements that did not violate the constraints (SI~section~S6). Contact triggered average plots show the snake's average behavior relative to when a new contact is formed (SI~section~S11). These were used to quantify force rearrangements relative to the nullspace and the fraction of rearrangement attributed to each body segment.

\section*{Author contributions}
C.A.R, J.R.M, and J.M.R designed research. C.A.R, M.L., Y.N., G.T., and J.M.R. performed research. C.A.R., M.L., and J.M.R. analyzed data. C.A.R., J.R.M, and J.M.R wrote the paper. The authors have no competing interests to declare.

\section*{Acknowledgments}
The authors thank Horace Dale and Lowell Ramsey for assistance in experimental design, and Jessica Tingle,  Jake Socha, and Gordon Berman for helpful discussions. C.A.R. was partially supported by NSF PHY-2310741.

\subsection*{Disclosure of Delegation to Generative AI} 

The authors declare the use of generative AI in the research and writing process. According to the GAIDeT taxonomy (2025), the following tasks were delegated to GAI tools under full human supervision: Code optimization, Proofreading and editing. The GAI tool used was: ChatGPT 5.5, Claude Opus 4.8. Responsibility for the final manuscript lies entirely with the authors.

\bibliographystyle{unsrt} 
\bibliography{refs}

@article{autumn2000adhesive,
  title={Adhesive force of a single gecko foot-hair},
  author={Autumn, Kellar and Liang, Yiching A and Hsieh, S Tonia and Zesch, Wolfgang and Chan, Wai Pang and Kenny, Thomas W and Fearing, Ronald and Full, Robert J},
  journal={Nature},
  volume={405},
  number={6787},
  pages={681--685},
  year={2000},
  publisher={Nature Publishing Group UK London}
}

@article{tytell_spikes_2011,
  author  = {Tytell, Eric D. and Holmes, Philip and Cohen, Avis H.},
  title   = {Spikes alone do not behavior make: why neuroscience needs biomechanics},
  journal = {Current Opinion in Neurobiology},
  volume  = {21}, number = {5}, pages = {816--822}, year = {2011},
  doi     = {10.1016/j.conb.2011.05.017}
}

@book{bernstein_coordination_1967,
  author    = {Bernstein, Nikolai A.},
  title     = {The Co-ordination and Regulation of Movements},
  publisher = {Pergamon Press}, address = {Oxford}, year = {1967}
}

@article{bicchi_hands_2000,
  author  = {Bicchi, Antonio},
  title   = {Hands for dexterous manipulation and robust grasping: a difficult road toward simplicity},
  journal = {IEEE Transactions on Robotics and Automation},
  volume  = {16}, number = {6}, pages = {652--662}, year = {2000},
  doi     = {10.1109/70.897777}
}

@article{daley_running_2006,
  author  = {Daley, Monica A. and Biewener, Andrew A.},
  title   = {Running over rough terrain reveals limb control for intrinsic stability},
  journal = {Proceedings of the National Academy of Sciences},
  volume  = {103}, number = {42}, pages = {15681--15686}, year = {2006},
  doi     = {10.1073/pnas.0601473103}
}

@article{scholz_uncontrolled_1999,
  author  = {Scholz, John P. and Sch{\"o}ner, Gregor},
  title   = {The uncontrolled manifold concept: identifying control variables for a functional task},
  journal = {Experimental Brain Research},
  volume  = {126}, number = {3}, pages = {289--306}, year = {1999},
  doi     = {10.1007/s002210050738}
}

@book{murray_mathematical_1994,
  author    = {Murray, Richard M. and Li, Zexiang and Sastry, S. Shankar},
  title     = {A Mathematical Introduction to Robotic Manipulation},
  publisher = {CRC Press}, address = {Boca Raton, FL}, year = {1994}
}

@article{kitano_biological_2004,
  author  = {Kitano, Hiroaki},
  title   = {Biological robustness},
  journal = {Nature Reviews Genetics},
  volume  = {5}, number = {11}, pages = {826--837}, year = {2004},
  doi     = {10.1038/nrg1471}
}

@article{carlson_complexity_2002,
  author  = {Carlson, J. M. and Doyle, John},
  title   = {Complexity and robustness},
  journal = {Proceedings of the National Academy of Sciences},
  volume  = {99}, number = {suppl 1}, pages = {2538--2545}, year = {2002},
  doi     = {10.1073/pnas.012582499}
}

@article{alexander_factors_1981,
  author  = {Alexander, R. McNeill},
  title   = {Factors of safety in the structure of animals},
  journal = {Science Progress},
  volume  = {67}, number = {266}, pages = {109--130}, year = {1981}
}

@article{pizzatto2007life,
	title={Life-history adaptations to arboreality in snakes},
	author={Pizzatto, L{\'\i}gia and Almeida-Santos, Selma M and Shine, Richard},
	journal={Ecology},
	volume={88},
	number={2},
	pages={359--366},
	year={2007},
	publisher={Wiley Online Library}
}

@article{rieser_geometric_2024,
	title = {Geometric phase predicts locomotion performance in undulating living systems across scales},
	volume = {121},
	issn = {0027-8424, 1091-6490},
	url = {https://pnas.org/doi/10.1073/pnas.2320517121},
	doi = {10.1073/pnas.2320517121},
	language = {en},
	number = {24},
	urldate = {2026-05-04},
	journal = {Proceedings of the National Academy of Sciences},
	author = {Rieser, Jennifer M. and Chong, Baxi and Gong, Chaohui and Astley, Henry C. and Schiebel, Perrin E. and Diaz, Kelimar and Pierce, Christopher J. and Lu, Hang and Hatton, Ross L. and Choset, Howie and Goldman, Daniel I.},
	month = jun,
	year = {2024},
	pages = {e2320517121},
}

@article{ijspeert_central_2008,
	title = {Central pattern generators for locomotion control in animals and robots: {A} review},
	volume = {21},
	copyright = {https://www.elsevier.com/tdm/userlicense/1.0/},
	issn = {08936080},
	shorttitle = {Central pattern generators for locomotion control in animals and robots},
	url = {https://linkinghub.elsevier.com/retrieve/pii/S0893608008000804},
	doi = {10.1016/j.neunet.2008.03.014},
	language = {en},
	number = {4},
	urldate = {2026-05-04},
	journal = {Neural Networks},
	author = {Ijspeert, Auke Jan},
	month = may,
	year = {2008},
	pages = {642--653},
}

@article{biswas_mode_2023,
	title = {Mode switching in organisms for solving explore-versus-exploit problems},
	volume = {5},
	issn = {2522-5839},
	url = {https://www.nature.com/articles/s42256-023-00745-y},
	doi = {10.1038/s42256-023-00745-y},
	language = {en},
	number = {11},
	urldate = {2026-05-04},
	journal = {Nature Machine Intelligence},
	author = {Biswas, Debojyoti and Lamperski, Andrew and Yang, Yu and Hoffman, Kathleen and Guckenheimer, John and Fortune, Eric S. and Cowan, Noah J.},
	month = oct,
	year = {2023},
	pages = {1285--1296},
}

@article{todorov_optimal_2002,
	title = {Optimal feedback control as a theory of motor coordination},
	volume = {5},
	copyright = {http://www.springer.com/tdm},
	issn = {1097-6256, 1546-1726},
	url = {https://www.nature.com/articles/nn963},
	doi = {10.1038/nn963},
	language = {en},
	number = {11},
	urldate = {2026-05-04},
	journal = {Nature Neuroscience},
	author = {Todorov, Emanuel and Jordan, Michael I.},
	month = nov,
	year = {2002},
	pages = {1226--1235},
}

@article{ramdya_neuromechanics_2023,
	title = {The neuromechanics of animal locomotion: {From} biology to robotics and back},
	volume = {8},
	issn = {2470-9476},
	shorttitle = {The neuromechanics of animal locomotion},
	url = {https://www.science.org/doi/10.1126/scirobotics.adg0279},
	doi = {10.1126/scirobotics.adg0279},
	language = {en},
	number = {78},
	urldate = {2026-05-04},
	journal = {Science Robotics},
	author = {Ramdya, Pavan and Ijspeert, Auke Jan},
	month = may,
	year = {2023},
	pages = {eadg0279},
}

@article{aguilar_review_2016,
	title = {A review on locomotion robophysics: the study of movement at the intersection of robotics, soft matter and dynamical systems},
	volume = {79},
	issn = {0034-4885, 1361-6633},
	shorttitle = {A review on locomotion robophysics},
	url = {https://iopscience.iop.org/article/10.1088/0034-4885/79/11/110001},
	doi = {10.1088/0034-4885/79/11/110001},
	language = {en},
	number = {11},
	urldate = {2026-05-04},
	journal = {Reports on Progress in Physics},
	author = {Aguilar, Jeffrey and Zhang, Tingnan and Qian, Feifei and Kingsbury, Mark and McInroe, Benjamin and Mazouchova, Nicole and Li, Chen and Maladen, Ryan and Gong, Chaohui and Travers, Matt and Hatton, Ross L and Choset, Howie and Umbanhowar, Paul B and Goldman, Daniel I},
	month = nov,
	year = {2016},
	pages = {110001},
}

@article{hoffmann2026postural,
  title={Postural control in an upright snake},
  author={Hoffmann, Ludwig A and Bryde, Petur and Davenport, Ian C and Prasath, S Ganga and Jayne, Bruce C and Mahadevan, L},
  journal={Journal of The Royal Society Interface},
  volume={23},
  number={235},
  year={2026},
  publisher={The Royal Society}
}

@article{jayne_surface_2012,
	title = {Surface shape affects the three-dimensional exploratory movements of nocturnal arboreal snakes},
	volume = {198},
	issn = {0340-7594, 1432-1351},
	url = {http://link.springer.com/10.1007/s00359-012-0761-y},
	doi = {10.1007/s00359-012-0761-y},
	language = {en},
	number = {12},
	urldate = {2021-08-26},
	journal = {Journal of Comparative Physiology A},
	author = {Jayne, Bruce C. and Olberding, Jeffrey P. and Athreya, Dilip and Riley, Michael A.},
	month = dec,
	year = {2012},
	pages = {905--913},
}

@article{jorgensen_three-dimensional_2017,
	title = {Three-dimensional trajectories affect the epaxial muscle activity of arboreal snakes crossing gaps},
	copyright = {http://www.biologists.com/user-licence-1-1/},
	issn = {1477-9145, 0022-0949},
	url = {https://journals.biologists.com/jeb/article/doi/10.1242/jeb.164640/262671/Three-dimensional-trajectories-affect-the-epaxial},
	doi = {10.1242/jeb.164640},
	language = {en},
	urldate = {2025-06-06},
	journal = {Journal of Experimental Biology},
	author = {Jorgensen, Ryan M. and Jayne, Bruce C.},
	month = jan,
	year = {2017},
	pages = {jeb.164640},
}

@article{tang_arboreal_2017,
	title = {Arboreal concertina locomotion of snake robots on cylinders},
	volume = {14},
	issn = {1729-8814, 1729-8814},
	url = {http://journals.sagepub.com/doi/10.1177/1729881417748440},
	doi = {10.1177/1729881417748440},
	language = {en},
	number = {6},
	urldate = {2021-09-17},
	journal = {International Journal of Advanced Robotic Systems},
	author = {Tang, Chaoquan and Shu, Xin and Meng, Deyuan and Zhou, Gongbo},
	month = nov,
	year = {2017},
	pages = {172988141774844},
}

@article{rollinson_pipe_2016,
	title = {Pipe {Network} {Locomotion} with a {Snake} {Robot}},
	volume = {33},
	copyright = {http://onlinelibrary.wiley.com/termsAndConditions\#vor},
	issn = {1556-4959, 1556-4967},
	url = {https://onlinelibrary.wiley.com/doi/10.1002/rob.21549},
	doi = {10.1002/rob.21549},
	language = {en},
	number = {3},
	urldate = {2026-04-13},
	journal = {Journal of Field Robotics},
	author = {Rollinson, David and Choset, Howie},
	month = may,
	year = {2016},
	pages = {322--336},
}

@article{jurestovsky_generation_2021,
	title = {Generation of propulsive force via vertical undulations in snakes},
	volume = {224},
	issn = {0022-0949, 1477-9145},
	url = {https://journals.biologists.com/jeb/article/224/13/jeb239020/270817/Generation-of-propulsive-force-via-vertical},
	doi = {10.1242/jeb.239020},
	language = {en},
	number = {13},
	urldate = {2021-09-03},
	journal = {Journal of Experimental Biology},
	author = {Jurestovsky, Derek J. and Usher, Logan R. and Astley, Henry C.},
	month = jul,
	year = {2021},
	pages = {jeb239020},
}

@article{marvi_friction_2012,
	title = {Friction enhancement in concertina locomotion of snakes},
	volume = {9},
	issn = {1742-5689, 1742-5662},
	url = {https://royalsocietypublishing.org/doi/10.1098/rsif.2012.0132},
	doi = {10.1098/rsif.2012.0132},
	language = {en},
	number = {76},
	urldate = {2021-06-15},
	journal = {Journal of The Royal Society Interface},
	author = {Marvi, Hamidreza and Hu, David L.},
	month = nov,
	year = {2012},
	pages = {3067--3080},
}

@article{wiens_why_2006,
	title = {Why does a trait evolve multiple times within a clade? Repeated evolution of snakeline body form in squamate reptiles},
	volume = {60},
	issn = {0014-3820, 1558-5646},
	shorttitle = {{WHY} {DOES} {A} {TRAIT} {EVOLVE} {MULTIPLE} {TIMES} {WITHIN} {A} {CLADE}?},
	url = {http://doi.wiley.com/10.1111/j.0014-3820.2006.tb01088.x},
	doi = {10.1111/j.0014-3820.2006.tb01088.x},
	language = {en},
	number = {1},
	urldate = {2021-05-16},
	journal = {Evolution},
	author = {Wiens, John J. and Brandley, Matthew C. and Reeder, Tod W.},
	month = jan,
	year = {2006},
	pages = {123--141},
}

@article{byrnes_gripping_2014,
	title = {Gripping during climbing of arboreal snakes may be safe but not economical},
	volume = {10},
	issn = {1744-9561, 1744-957X},
	url = {https://royalsocietypublishing.org/doi/10.1098/rsbl.2014.0434},
	doi = {10.1098/rsbl.2014.0434},
	language = {en},
	number = {8},
	urldate = {2021-08-26},
	journal = {Biology Letters},
	author = {Byrnes, Greg and Jayne, Bruce C.},
	month = aug,
	year = {2014},
	pages = {20140434},
}

@article{jayne_what_2020,
	title = {What {Defines} {Different} {Modes} of {Snake} {Locomotion}?},
	volume = {60},
	issn = {1540-7063, 1557-7023},
	url = {https://academic.oup.com/icb/article/60/1/156/5818495},
	doi = {10.1093/icb/icaa017},
	language = {en},
	number = {1},
	urldate = {2021-08-26},
	journal = {Integrative and Comparative Biology},
	author = {Jayne, Bruce C},
	month = jul,
	year = {2020},
	pages = {156--170},
}

@article{astley_arboreal_2009,
	title = {Arboreal habitat structure affects the performance and modes of locomotion of corn snakes (\textit{{Elaphe} guttata})},
	volume = {311A},
	issn = {19325223, 19325231},
	url = {https://onlinelibrary.wiley.com/doi/10.1002/jez.521},
	doi = {10.1002/jez.521},
	language = {en},
	number = {3},
	urldate = {2021-08-29},
	journal = {Journal of Experimental Zoology Part A: Ecological Genetics and Physiology},
	author = {Astley, Henry C. and Jayne, Bruce C.},
	month = mar,
	year = {2009},
	pages = {207--216},
}

@article{jayne_why_2015,
	title = {Why arboreal snakes should not be cylindrical: body shape, incline and surface roughness have interactive effects on locomotion},
	volume = {218},
	issn = {1477-9145, 0022-0949},
	shorttitle = {Why arboreal snakes should not be cylindrical},
	url = {https://journals.biologists.com/jeb/article/218/24/3978/14344/Why-arboreal-snakes-should-not-be-cylindrical-body},
	doi = {10.1242/jeb.129379},
	language = {en},
	number = {24},
	urldate = {2021-08-29},
	journal = {Journal of Experimental Biology},
	author = {Jayne, Bruce C. and Newman, Steven J. and Zentkovich, Michele M. and Berns, H. Matthew},
	month = dec,
	year = {2015},
	pages = {3978--3986},
}

@article{astley_effects_2007,
	title = {Effects of perch diameter and incline on the kinematics, performance and modes of arboreal locomotion of corn snakes ({Elaphe} guttata)},
	volume = {210},
	language = {en},
	journal = {Journal of Experimental Biology},
	author = {Astley, Henry C and Jayne, Bruce C},
	year = {2007},
	pages = {3862--3872},
}

@article{hunt_acrobatic_2021,
	title = {Acrobatic squirrels learn to leap and land on tree branches without falling},
	volume = {373},
	issn = {0036-8075, 1095-9203},
	url = {https://www.sciencemag.org/lookup/doi/10.1126/science.abe5753},
	doi = {10.1126/science.abe5753},
	language = {en},
	number = {6555},
	urldate = {2021-09-07},
	journal = {Science},
	author = {Hunt, Nathaniel H. and Jinn, Judy and Jacobs, Lucia F. and Full, Robert J.},
	month = aug,
	year = {2021},
	pages = {697--700},
}

@article{cant_positional_1992,
	title = {Positional behavior and body size of arboreal primates: {A} theoretical framework for field studies and an illustration of its application},
	volume = {88},
	issn = {0002-9483, 1096-8644},
	shorttitle = {Positional behavior and body size of arboreal primates},
	url = {https://onlinelibrary.wiley.com/doi/10.1002/ajpa.1330880302},
	doi = {10.1002/ajpa.1330880302},
	language = {en},
	number = {3},
	urldate = {2022-02-22},
	journal = {American Journal of Physical Anthropology},
	author = {Cant, John G. H.},
	month = jul,
	year = {1992},
	pages = {273--283},
}

@article{jayne1986kinematics,
  title={Kinematics of terrestrial snake locomotion},
  author={Jayne, Bruce C},
  journal={Copeia},
  pages={915--927},
  year={1986},
  publisher={JSTOR}
}

@book{lillywhite2014snakes,
  title={How snakes work: structure, function and behavior of the world's snakes},
  author={Lillywhite, Harvey B},
  year={2014},
  publisher={Oxford University Press}
}

@incollection{Gasc1981,
  author    = {Jean-Pierre Gasc},
  title     = {Axial musculature},
  booktitle = {Biology of the Reptilia},
  editor     = {Carl Gans and Thomas S. Parsons},
  pages      = {355--435},
  year       = {1981},
  publisher  = {Academic Press},
  address    = {London}
}

@article{tingle2024functional,
  title={Functional diversity of snake locomotor behaviors: a review of the biological literature for bioinspiration},
  author={Tingle, Jessica L and Garner, Kelsey L and Astley, Henry C},
  journal={Annals of the New York Academy of Sciences},
  volume={1533},
  number={1},
  pages={16--37},
  year={2024},
  publisher={Wiley Online Library}
}

@article{fischer_evolution_2010,
	title = {Evolution of chameleon locomotion, or how to become arboreal as a reptile},
	volume = {113},
	issn = {09442006},
	url = {https://linkinghub.elsevier.com/retrieve/pii/S0944200609000543},
	doi = {10.1016/j.zool.2009.07.001},
	language = {en},
	number = {2},
	urldate = {2022-02-22},
	journal = {Zoology},
	author = {Fischer, Martin S. and Krause, Cornelia and Lilje, Karin E.},
	month = mar,
	year = {2010},
	pages = {67--74},
}

@article{jusufi_active_2008,
	title = {Active tails enhance arboreal acrobatics in geckos},
	volume = {105},
	issn = {0027-8424, 1091-6490},
	url = {http://www.pnas.org/cgi/doi/10.1073/pnas.0711944105},
	doi = {10.1073/pnas.0711944105},
	language = {en},
	number = {11},
	urldate = {2022-02-22},
	journal = {Proceedings of the National Academy of Sciences},
	author = {Jusufi, A. and Goldman, D. I. and Revzen, S. and Full, R. J.},
	month = mar,
	year = {2008},
	pages = {4215--4219},
}

@article{savidge_lasso_2021,
	title = {Lasso locomotion expands the climbing repertoire of snakes},
	volume = {31},
	issn = {09609822},
	url = {https://linkinghub.elsevier.com/retrieve/pii/S0960982220317632},
	doi = {10.1016/j.cub.2020.11.050},
	language = {en},
	number = {1},
	urldate = {2022-02-24},
	journal = {Current Biology},
	author = {Savidge, Julie A. and Seibert, Thomas F. and Kastner, Martin and Jayne, Bruce C.},
	month = jan,
	year = {2021},
	pages = {R7--R8},
}

@article{tingle_relative_2023,
	title = {The relative contributions of multiarticular snake muscles to movement in different planes},
	volume = {284},
	issn = {0362-2525, 1097-4687},
	url = {https://onlinelibrary.wiley.com/doi/10.1002/jmor.21591},
	doi = {10.1002/jmor.21591},
	language = {en},
	number = {6},
	urldate = {2023-07-24},
	journal = {Journal of Morphology},
	author = {Tingle, Jessica L. and Jurestovsky, Derek J. and Astley, Henry C.},
	month = jun,
	year = {2023},
	pages = {e21591},
}

@article{jayne_muscular_1988,
	title = {Muscular mechanisms of snake locomotion: {An} electromyographic study of lateral undulation of the florida banded water snake (\textit{{Nerodia} fasciata}) and the yellow rat snake (\textit{{Elaphe} obsoleta})},
	volume = {197},
	issn = {0362-2525, 1097-4687},
	shorttitle = {Muscular mechanisms of snake locomotion},
	url = {https://onlinelibrary.wiley.com/doi/10.1002/jmor.1051970204},
	doi = {10.1002/jmor.1051970204},
	language = {en},
	number = {2},
	urldate = {2023-10-26},
	journal = {Journal of Morphology},
	author = {Jayne, Bruce C.},
	month = aug,
	year = {1988},
	pages = {159--181},
}

@article{jayne_muscular_1988-1,
	title = {Muscular {Mechanisms} of {Snake} {Locomotion}: an {Electromyographic} {Study} of the {Sidewinding} and {Concertina} {Modes} of \textit{{Crotalus} {Cerastes}, {Nerodia} {Fasciata}} and \textit{{Elaphe} {Obsoleta}}},
	volume = {140},
	copyright = {http://www.biologists.com/user-licence-1-1/},
	issn = {0022-0949, 1477-9145},
	shorttitle = {Muscular {Mechanisms} of {Snake} {Locomotion}},
	url = {https://journals.biologists.com/jeb/article/140/1/1/5756/Muscular-Mechanisms-of-Snake-Locomotion-an},
	doi = {10.1242/jeb.140.1.1},
	language = {en},
	number = {1},
	urldate = {2024-10-01},
	journal = {Journal of Experimental Biology},
	author = {Jayne, Bruce C.},
	month = nov,
	year = {1988},
	pages = {1--33},
}

@article{dirks_fluid-based_2011,
	title = {Fluid-based adhesion in insects – principles and challenges},
	volume = {7},
	issn = {1744-683X, 1744-6848},
	url = {https://xlink.rsc.org/?DOI=c1sm06269g},
	doi = {10.1039/c1sm06269g},
	language = {en},
	number = {23},
	urldate = {2025-03-26},
	journal = {Soft Matter},
	author = {Dirks, Jan-Henning and Federle, Walter},
	year = {2011},
	pages = {11047},
}

@article{labonte_extreme_2016,
	title = {Extreme positive allometry of animal adhesive pads and the size limits of adhesion-based climbing},
	volume = {113},
	issn = {0027-8424, 1091-6490},
	url = {https://pnas.org/doi/full/10.1073/pnas.1519459113},
	doi = {10.1073/pnas.1519459113},
	language = {en},
	number = {5},
	urldate = {2025-03-26},
	journal = {Proceedings of the National Academy of Sciences},
	author = {Labonte, David and Clemente, Christofer J. and Dittrich, Alex and Kuo, Chi-Yun and Crosby, Alfred J. and Irschick, Duncan J. and Federle, Walter},
	month = feb,
	year = {2016},
	pages = {1297--1302},
}

\onecolumn

\beginsupplement
\newpage
\setcounter{page}{1}

\section*{Supplemental Information}

\noindent Movies can be downloaded from: 
\href{https://osf.io/x5rdu/overview?view_only=3a8c024555c747d0a15d5f2bab60ec9e}{https://osf.io/x5rdu}. 

\noindent\textbf{Supplemental movie S1.}\\
Upward climbing with $2.5$--cm posts spaced $50$~cm apart. In all of these movies, the black and white video on the left is a video of the experiment where we track reflective markers placed on the back of the snake. The animation on the right shows the tracked body of the snake and the arrows represent the force on each post. Dotted lines show a path across the post perpendicular to the snake's contact location which makes the deflection apparent. Here forces are generally deflected below this line indicating a tangential force opposite what is predicted by friction. Video is 5x speed.
\\
\textbf{Supplemental movie S2.}\\
Downward climbing with $2.5$--cm posts spaced $50$~cm apart. Video is 5x speed.
\\
\textbf{Supplemental movie S3.}\\
Upward climbing with $2.5$--cm posts spaced $100$~cm apart. Video is 5x speed.
\\
\textbf{Supplemental movie S4.}\\
Downward climbing with $2.5$--cm posts spaced $100$~cm apart. Video is 5x speed.
\\
\textbf{Supplemental movie S5.}\\
Upward climbing with $0.3$--cm posts spaced $50$~cm apart. Video is 5x speed.
\\
\textbf{Supplemental movie S6.}\\
Downward climbing with $0.3$--cm posts spaced $50$~cm apart. Video is 5x speed.
\\
\textbf{Supplemental movie S7.}\\
Upward climbing from our robotic model. The force sensitive post is colored cyan in the animation. Video is 5x speed.
\\
\textbf{Supplemental movie S8.}\\
Downward climbing from our robotic model. Video is 5x speed.
\\
\textbf{Supplemental movie S9.}\\
Highlighting an instance in movie S5 where the middle of the body changes sides of the post demonstrating active readjustment in the mid-body. Video is 5x speed.
\\
\textbf{Supplemental movie S10.}\\
Brooks' Kingnakes (\textit{Lampropeltis getula brooksi} perform poorly and fail on our climbing wall. Clips are real speed.\\

\section{Acrylic surface with no posts}

\begin{figure}
    \centering
    \includegraphics[width=\textwidth]{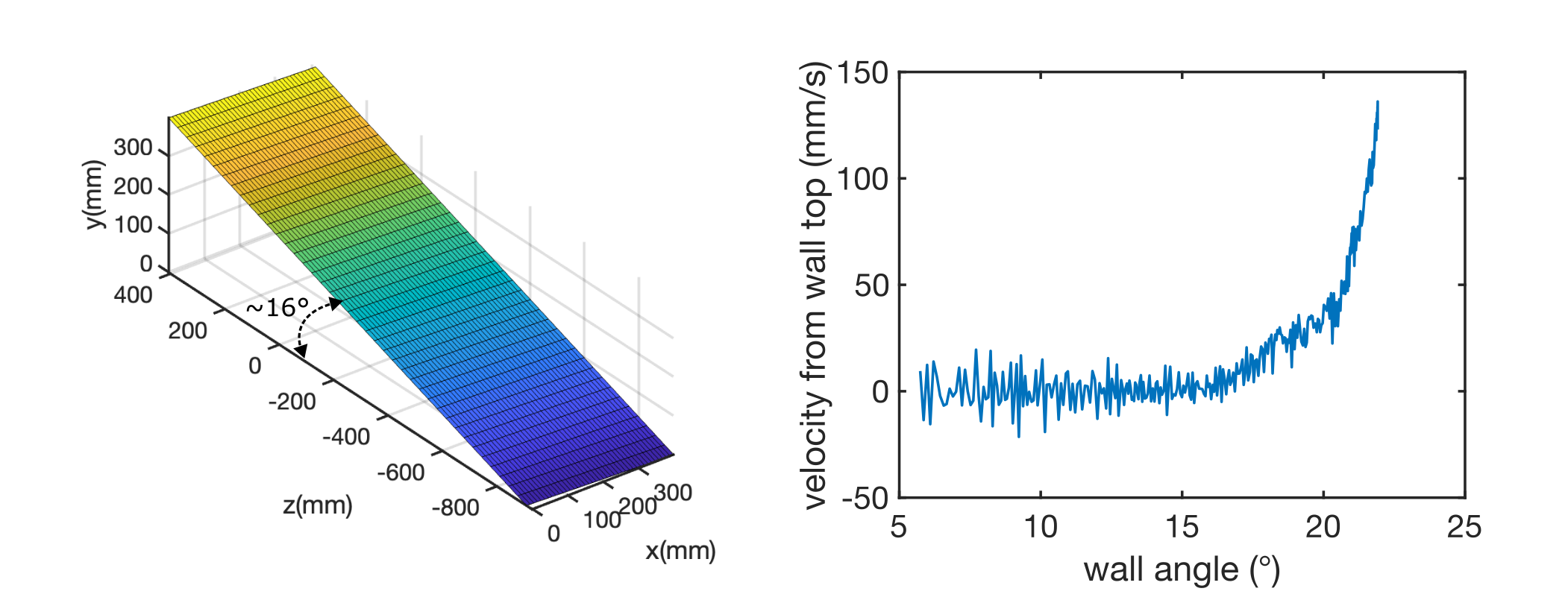}
    \caption{\textbf{Wall tilting.} Here we tilted a flat acrylic wall with no posts by hand until it reached an angle (left) where the snake started sliding indicated by its velocity relative to the end of the wall (right).}
    \label{fig:tilt}
\end{figure}

We performed 12 experiments with 2 individuals (ta and ch listed above) to gauge the greatest incline that would be passible for a snake without dedicated surface features on the acrylic wall. We did this by starting with the wall level (parallel to the ground) and placed the snake on one end. We then elevated that end of the wall by hand until the snake slid down. We placed two reflective markers on the snake, one on the head and one on the mid-body, and used our marker tracking setup to track the snake's position over time. Four markers were placed on each corner of the wall and remained visible through the trial. This allowed us to fit a plane to these markers and calculate the incline angle of the wall at each point in time. Figure~\ref{fig:tilt} shows, for a single trial, the snake's velocity relative to the wall end based on the angle of the wall. For these twelve trials, the incline angle where snakes started to slide was $16\pm 2^{\circ}$ demonstrating the impossibility of snakes traversing this surface above this incline without post contacts.

\pagebreak

\section{Trial counts by individual}

\begin{table}[h!]
    \centering
    \begin{tabular}{c|c|c}
        Animal ID & mass (g) & length (cm) \\
        \hline 
         ta & 96--103 & 74.4--75.4 \\
         ch & 101--112 & 79.0--85.1 \\
         el & 99--120 & 71.9--77.5 \\
         ca & 80--95 & 73.4--75.2 \\
         po & 87--104 & 70.1--73.4\\
    \end{tabular}
    \caption{Masses and lengths of the snakes during experimental phase.}
    \label{tab:table1}
\end{table}

\begin{table}[h!]
    \centering
    \begin{tabular}{c|c|c|c|c}
        Animal ID & 25/50 & 3/50 & 25/100 & total\\
        \hline 
         ta & 10 & 10 & 8 & 28\\
         ch & 11 & 10 & 8 & 29\\
         el & 10 & 8 &  7 & 25\\
         ca & 7 & 9 & 7 & 23\\
         po & 10 & 2 & 9 & 21\\
         \hline
         total & 48 & 39 & 39 & 126
    \end{tabular}
    \caption{Number of upward climbs for conditions labeled as post length (mm)/post spacing (mm).}
    \label{tab:table2}
\end{table}

\begin{table}[h!]
    \centering
    \begin{tabular}{c|c|c|c|c}
        Animal ID & 25/50 & 3/50 & 25/100 & total\\
        \hline 
         ta & 9 & 6 & 3 & 18\\
         ch & 11 & 5 & 9 & 25\\
         el & 8 & 10 & 5 & 23\\
         ca & 10 & 1 & 6 & 17\\
         po & 10 & 2 & 6 & 18\\
         \hline
         total & 48 & 24 & 29 & 101
    \end{tabular}
    \caption{Number of downward climbs for conditions labeled as post length (mm)/post spacing (mm).}
    \label{tab:table3}
\end{table}

\pagebreak

\section{Bracing force}
With shorter posts, forces were directed more laterally creating a high bracing effect. While still supporting their weight with an increased number of contacts, lateral forces increased by $\sim1$ body weight in total magnitude.

\begin{figure}
    \centering
    \includegraphics[width=\textwidth]{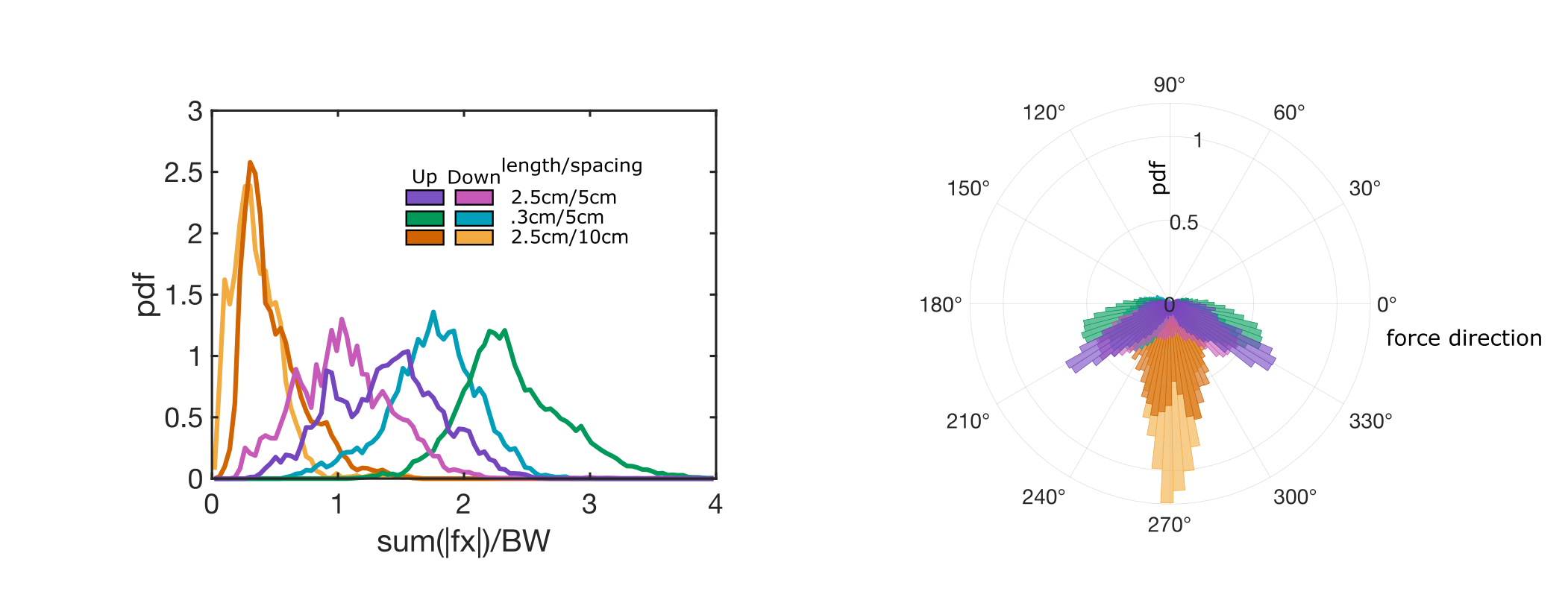}
    \caption{\textbf{Bracing force.} (left) The total magnitude of horizontal force at every point in time. (right) a polar histogram showing the direction of force applied to each post.}
    \label{fig:brace}
\end{figure}

\begin{figure}
    \centering
    \includegraphics[width=\textwidth]{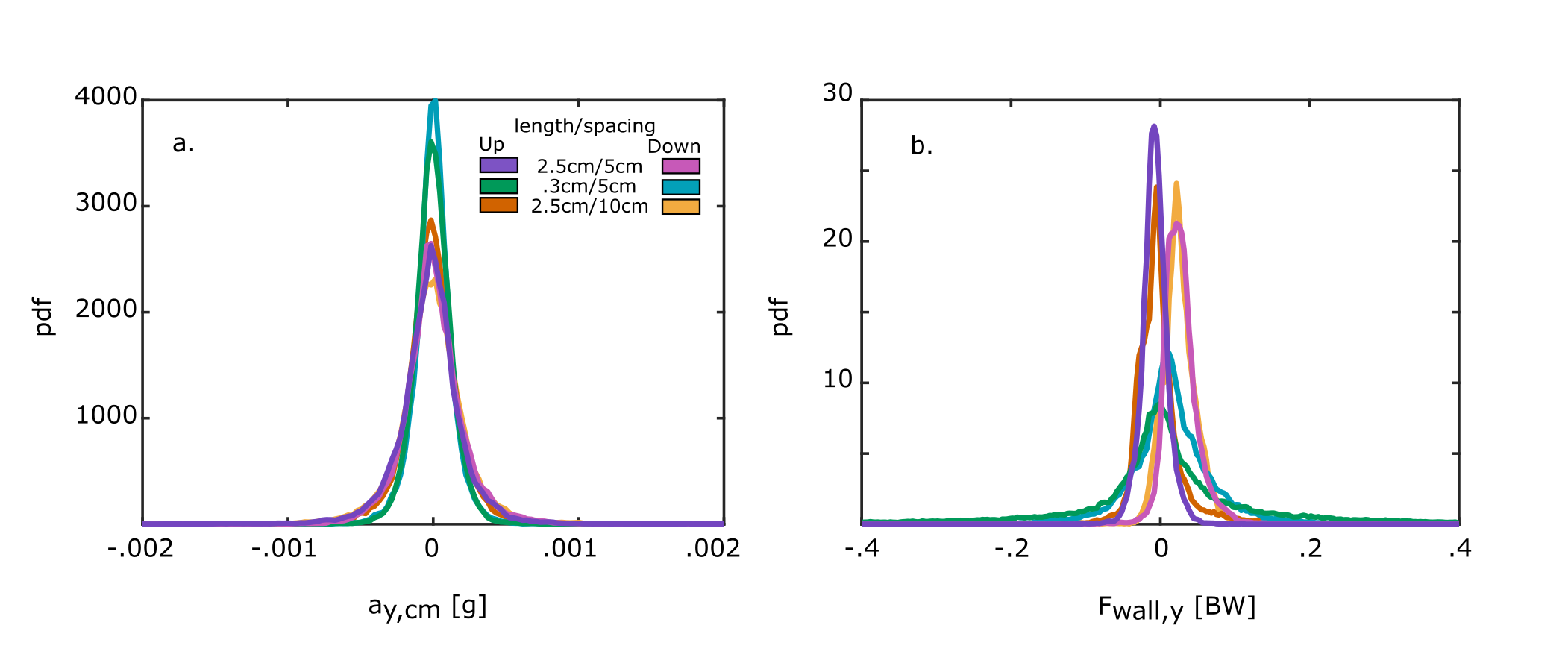}
    \caption{\textbf{Accelerations.} (left) The acceleration of the snake's body in each case was either small or zero. (right) Any residual force in the y-direction can be assumed to be from the snake slightly using the wall itself for support.}
    \label{fig:brace}
\end{figure}

\section{Forces in the third dimension}
A major limitation of our setup is that we can only measure forces in 2D in the plane of the wall. Adding a third dimension of force resolution would be mechanically costly for the setup, but we sought to get an estimation for whether the snakes are pulling on the posts to generate friction with them and the wall itself. To do this, we replaced the force sensors attached to the 12th post from the bottom with a 6-axis load cell. With this we were able to detect forces directed into and out of the wall at a single point. 

We then completed between 3 and 6 trials each with 2.5~cm posts and 0.3~cm posts spaced 5~cm apart; only two individuals (el and ch) were employed for these trials. We found the amount of force that the snakes pull on the posts with not insignificant (Fig.~\ref{fig:3D}). With longer posts, the snake consistently pulled on the post with a force about $1\%$ of its body weight. When the posts were shortened, however, and generating friction became more difficult, the force directed outward on the post increased to over $5\%$ in some instances. This would indicate that there is a friction force directed into the wall on the snake keeping it from sliding outward and cantilevering off the wall. 

\begin{figure}
    \centering
    \includegraphics[width=\textwidth]{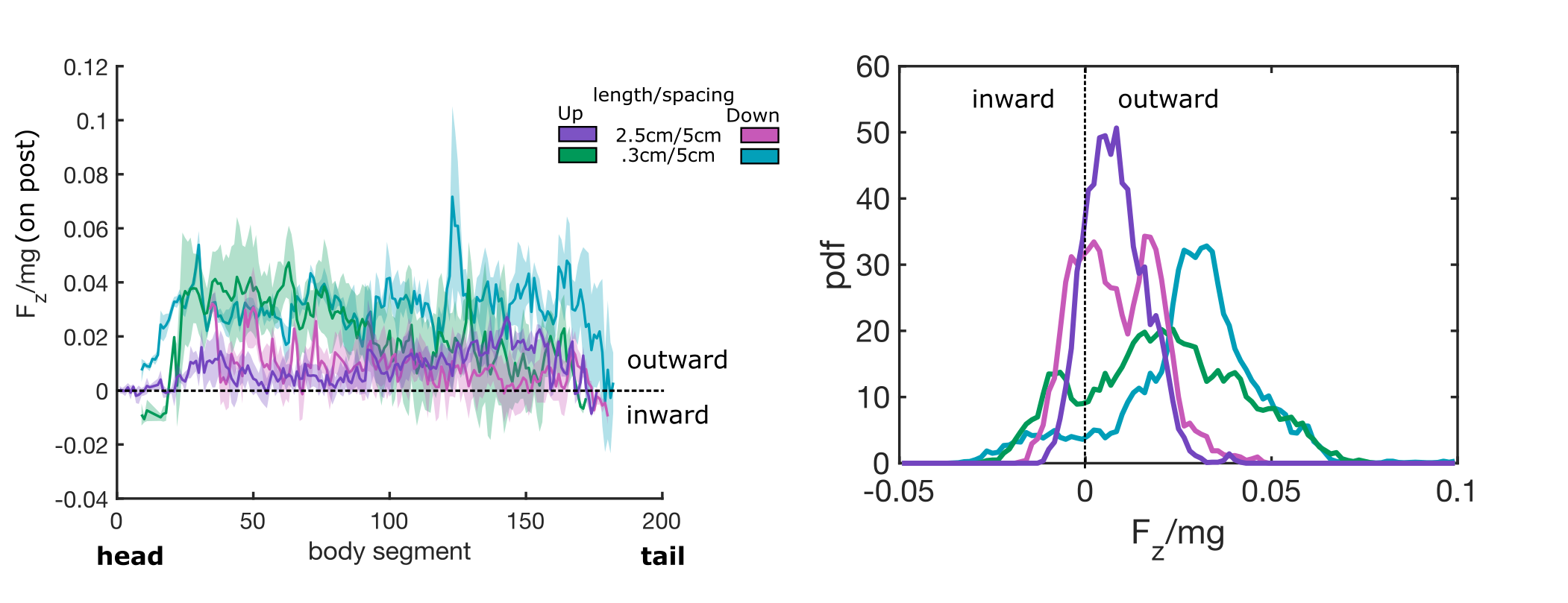}
    \caption{\textbf{Third dimension forces directed into and out of the wall.} A six axis load cell allowed us to track if the snakes were pushing or pulling on a single post perpendicular to the wall plane. The left plot shows the average amount of force applied to this post as each body segment slides past. The distribution of pulling force is even along the body except for the very ends. The right plot shows a histogram of all of the third-dimension forces applied to the post which inflate directed outward for shorter posts.}
    \label{fig:3D}
\end{figure}

\begin{figure}
    \centering
    \includegraphics[width=\textwidth]{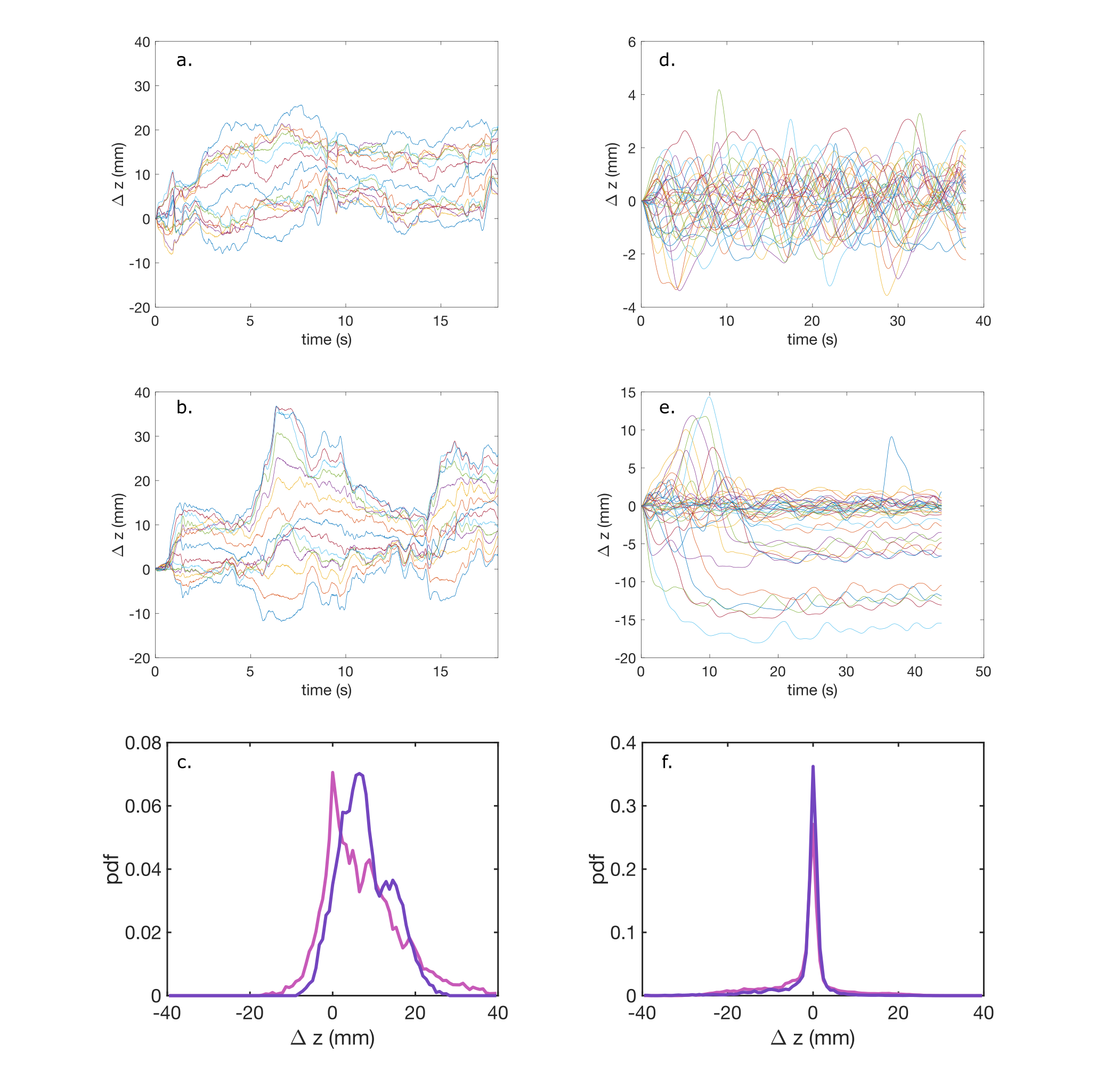}
    \caption{\textbf{Third dimension motion.} a-c robot, d-f snake}
    \label{fig:3D}
\end{figure}

\section{Metrics relative to contact location}
\begin{figure}
    \centering
    \includegraphics[width=\textwidth]{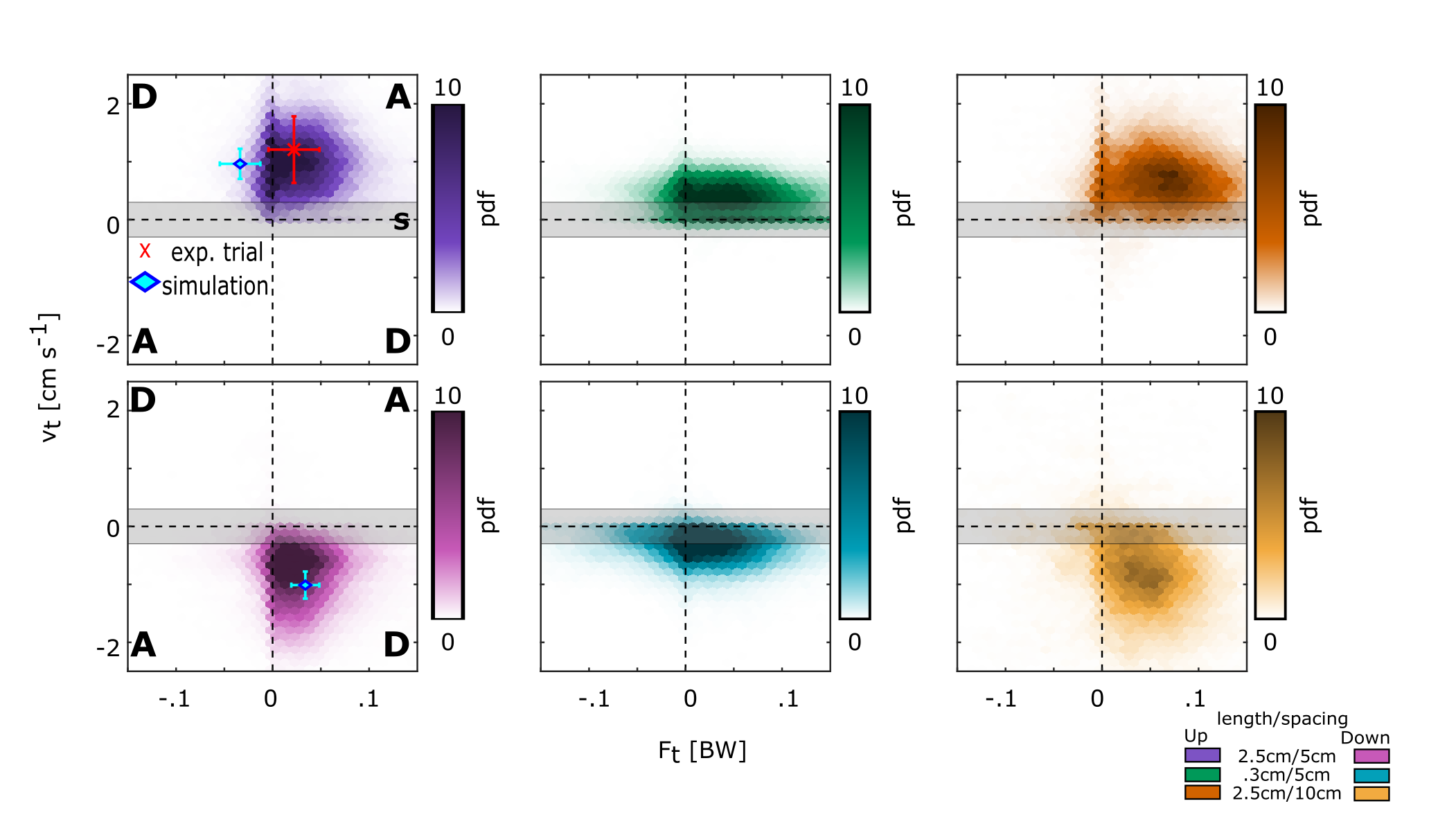}
    \caption{\textbf{Heatmaps of $F_t$ and $v_t$ for each post configuration.} Upward climbs in all conditions involve mainly active contacts where downward climbs remain dissipative. There are, however, an increase in static contacts when posts are shortened which is characteristic of the concertina gait.}
    \label{fig:powerAll}
\end{figure}

To detect individual contacts in our data, the splined snake body must come within a contact threshold of $17$~mm and apply a force comfortably outside the noise range of our sensors ($0.21$~g). The radial contact locations $\phi$ and the velocity at each contact using a weighted average of all of the segments within the contact threshold 
\begin{gather*}
    \phi=\sum_{seg}w_{seg}*\phi_{seg}/N_{seg}\\
    \vec{v}=\sum_{seg}w_{seg}*\vec{v}_{seg}/N_{seg}
\end{gather*}
where the weights $w_{seg}$ are assigned as the distance from the segment to the contact threshold boundary and normalized at each contact and $N_{seg}$ is the number of segments at the contact. $\phi$ was used to break the velocity and force at each contact into their tangential and normal components. Following from this, the tangential velocity was 
$$
v_t=\sqrt{\|\vec{v}\|-(\vec{v}\cdot \hat{\phi})}
$$
where $\hat{\phi}$ is a unit vector pointing from the center of the post to the edge at angle $\phi$ and the tangential force was expressed as 
$$ 
F_t=\|\vec{F}\|*sin(\alpha)
$$.
where $\alpha$, the force deflection angle, is defined as the difference between $\vec{F}$'s direction and an angle $180^\circ$ from $\phi$. We could then perform the contact analysis displayed in Figure 4 establishing each as active or dissipative. Similar plots for our other conditions can be seen in Figure~\ref{fig:powerAll}.

\begin{figure}
    \centering
    \includegraphics[width=0.5\textwidth]{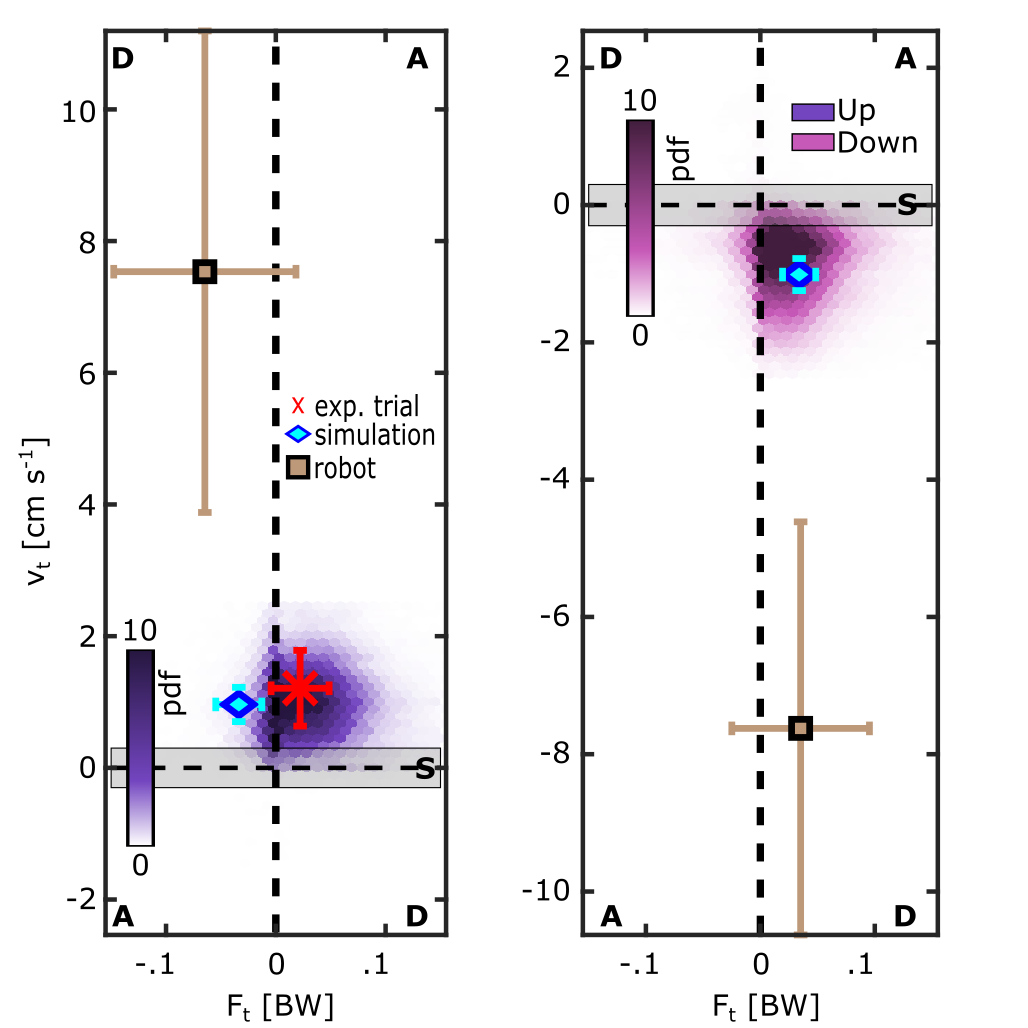}
    \caption{\textbf{Heatmaps of $F_t$ and $v_t$ as in Figure~4 including robot measurements.} Contact interactions in robot experiments tend to be dissipative as with the computational model.}
    \label{fig:powerwrobot}
\end{figure}

\section{Nullspace calculation}

Each force configuration can be represented by the horizontal and vertical force on each post $\vec{f_0}=[f_{x1},f_{y1},f_{x2},f_{y2},\dots, f_{xi},f_{yi}, \dots, f_{xN_p},f_{yN_p}]^T$ where $i$ denotes the post index and $N_p$ is the total number of posts on the wall. The balance space $\textbf{B}$ can be represented by a matrix 
$$\textbf{B}=
\begin{bmatrix}
    1 & 0 & 1 & 0 & \dots & 1 & 0 \\
    0 & 1 & 0 & 1 & \dots & 0 & 1\\ 
    -(y_1-y_{cm}) & x_1-x_{cm} & -(y_2-y_{cm}) & x_2-x_{cm} & \dots & -(y_{N_p}-y_{cm}) & x_{N_p}-x_{cm}
\end{bmatrix}
$$
which when multiplied by a valid $\vec{f_0}$ within the space gives $\textbf{B}\vec{f_0}=\vec{c}$ where $\vec{c}=[0, Mg, 0]^T$ reflecting the constraints 
\begin{gather*}
    \sum F_x=0 \\
    \sum F_y=Mg \\
    \sum \tau=0
\end{gather*}

As the snake establishes new contacts, each change in the force rearrangements must still satisfy the constraints such that $\textbf{B}(\vec{f_0}+\Delta \vec{f})=\vec{c}$, thus we can say that the rearrangement $\Delta \vec{f}$ lies within the ``nullspace'' of $\textbf{B}$.

\section{Bootstrapping fit parameters}
To find the bounds on the fit parameters for the nullspce function, we used hierarchical bootstrapping. We first resampled the data with replacement for individuals picking $5$ each time. We then resampled the trials for each individual, and then resampled events (new post contacts) within each trial. We put bounds on the bootstrapping so that the same number of events was chosen each time. We bootstrapped and fit the nullspace trajectory for 5000 iterations to establish a range for the fit parameters presented.

\section{Measuring snake length}
Without anesthetizing a snake, it is difficult to measure its length directly. Our goal was for these experiments to be as non-invasive as possible so anesthesia was not used. Instead, we measured snake lengths digitally. We first took a picture from directly overhead of the snake resting on a white surface next to a scale bar so that we could assign length values to the image pixels. We then binarized the images in MATLAB, dilated, and eroded the images in order to isolate a silhouette of the snake. We then used a built in skeletonization algorithm to calculate the length of the snake in pixels which was translated to real units with our scale bar.

\section{Force sensor calibration}

Each force-sensing post on the climbing wall consists of two $500$-g capacity load cells. Each load cell contains strain gauges wired in a bridge circuit so that changes in resistance of the strain gauges cause a voltage difference across the bridge corresponding to the applied load. Each load cell is connected to its own HX711 analog-to-digital conversion chip. The digital signals are then read using an Arduino mega. 

\begin{figure}
    \centering
    \includegraphics[width=\textwidth]{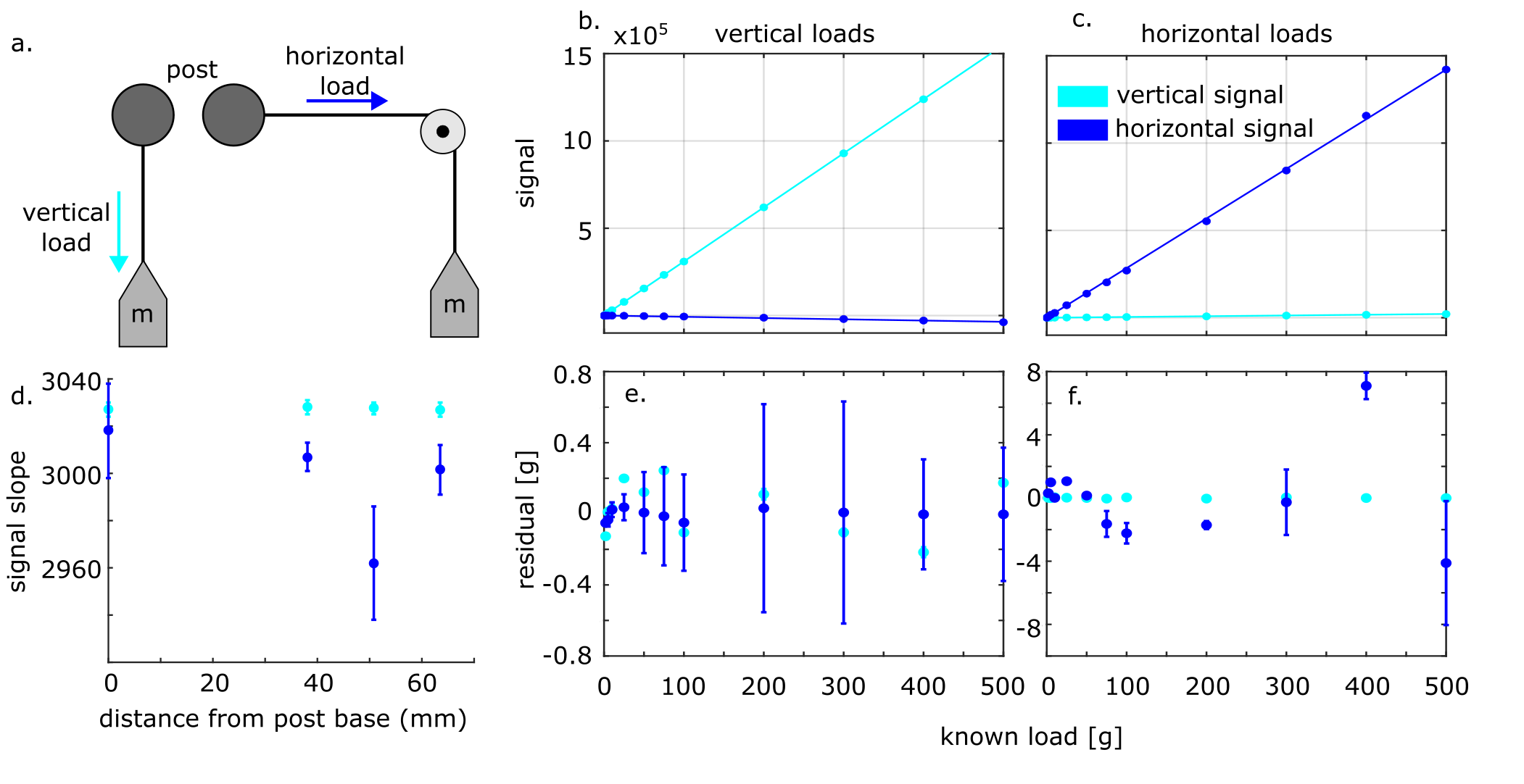}
    \caption{\textbf{Force sensor calibration} \textbf{(a)} A series of known weights were applied to each force sensor both vertically and horizontally to calibrate the two load cells. Vertical loads \textbf{(b)} and horizontal loads \textbf{(c)} caused a large linear signal in the load cell oriented in each direction and a small cross-signal in the other load cell. \textbf{(d)} The slope of the signal did not depend on the location on the post that the load was placed. After calibration, residuals were small for vertical signals and slightly larger for horizontal signals for both vertical \textbf{(e)} and horizontal \textbf{(f)} loads.}
    \label{fig:calibration}
\end{figure}

Each load cell has to be calibrated so that forces can be determined from the raw signals. The raw signals are linearly proportional to the applied load. Independently, the load cells can be calibrated by hanging a series of known masses from the end of the load cell and calculating the slope of the signal produced for each load. We performed a similar series of calibrations for each of our load cells. To calibrate the vertically oriented load cells, we hung weights of 0g, 2g, 5g, 10g, 25g, 50g, 75g, 100g, 200g, 300g, 400g, and 500g from the attached post and recorded the raw signal from each for 10 seconds (Figure \ref{fig:calibration}a). The average of the signal over 10s could be plotted for each load to determine the linear calibration factor that should be used for each (Figure \ref{fig:calibration}b). For horizontally oriented load cells, we used the same sequence of weights, but used a pulley system to load the post directly horizontally (Figure \ref{fig:calibration}a). A series of experiments loading the post at different distances determined that the signal produced was independent of where on the post the load was placed (Figure \ref{fig:calibration}d). All weights in this calibration procedure were loaded $12.7$~mm from the end of the post. 

The way our load cells were arranged, the signals from each pair ended up being coupled, i.e. a purely vertical load produced a signal read by the horizontally oriented load cell and vice versa (Figure \ref{fig:calibration}b,c).). Therefore, we used a linear system of the cross-signals to reproduce the known load and calculate a calibration matrix to determine measured forces in our experiments from the signals produced by each load cell pair: 
\begin{gather}
    \overrightarrow{F} =\textbf{B}*\overrightarrow{V}\\
    \begin{pmatrix}
        F_x\\
        F_y
    \end{pmatrix}
    = \textbf{B} * 
    \begin{pmatrix}
        V_x\\
        V_y\\
        V_x^2\\
        V_y^2
    \end{pmatrix}
    \label{eq:calibration}
\end{gather}

where $\textbf{B}$ is the $2\times4$ calibration matrix, $F_x$ and $F_y$ are the known loads applied horizontally and vertically respectively, and $V_x$ and $V_y$ are the raws signals produced by each load cell ($x$ corresponding to the horizontally oriented one, $y$ corresponding to vertical). Adding quadratic terms to our fitting reduced residuals compared to using linear terms only (Figure \ref{fig:calibration}d). 

\section{Randomizing trials}
Though versatile in the environmental configurations it can replicate, our experimental apparatus requires some level of reconstructing in order to change. We therefore could not randomize environmental conditions as that would require rotating between them. Instead, we completed all of the trials for one condition before moving to the next.

To introduce as much randomness as possible to account for trial ordering, we selected the individuals and climb directions for each trial at random. Each day, 2--3 snakes were made available for trials. For each trial, each snake was assigned a number and a random number generator in MATLAB would select which individual got used. We also had this random number generator choose whether the snake is to climb up or down the wall. We continued with this process until snakes reached their trial maxima for the day. If one snake completed its quota before the others, it was removed from the random pool and trials continued randomly with the remaining snakes.

\section{Contact triggered average plots}
Each time the snake established contact with a new post meeting our distance and force thresholds, we marked it as an event. We then shifted force data on the new post and each of the posts contacting further down the body relative to the start of this event. The average signal on each contact could then be tracked over time. We found that across trials, after a brief period of rearrangement, the force magnitude on each post ramped up to a stable, characteristic value. This value was slightly higher and more variable with shorter posts but still stereotyped across trials. With a larger spacing there were fewer contacts to work with meaning a greater force had to be applied ot each to support the snake's weight.

\begin{figure}
    \centering
    \includegraphics[width=\textwidth]{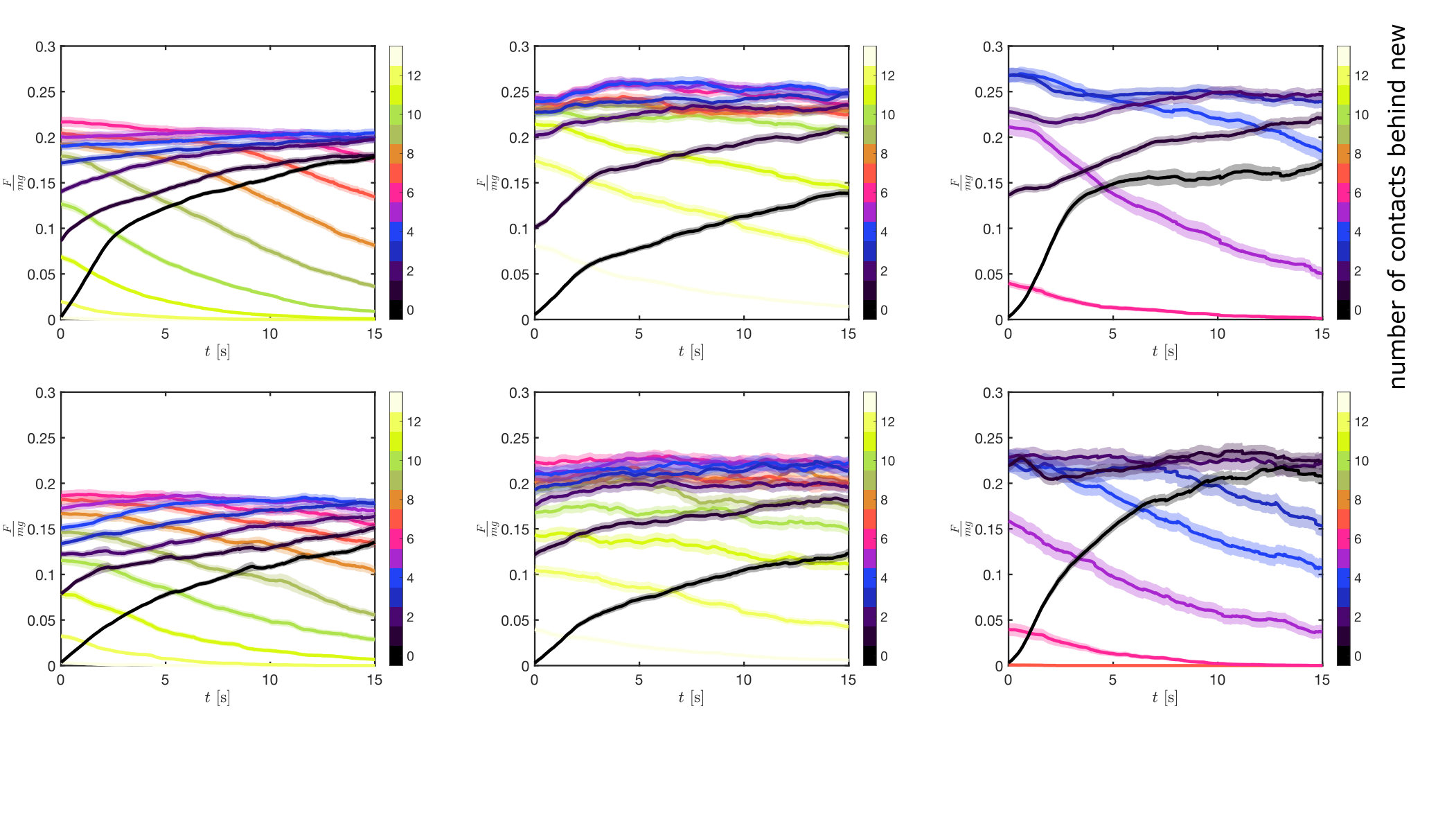}
    \caption{\textbf{Contact triggered average plots for the force magnitude on each post.} Shifting the data relative to when a new contact is formed allows us to see what the rest of the body does to accommodate this new contact.}
    \label{fig:CTA}
\end{figure}

\end{document}